\documentclass[12pt]{article}
\pdfoutput=1
\usepackage{graphicx}
\usepackage{amsmath}
\usepackage{amssymb}
\usepackage{amscd}
\usepackage{amsthm}
\usepackage{amssymb}
\usepackage{fullpage}
\usepackage{setspace}
\usepackage[hang,small]{caption}
\usepackage{authblk}
\usepackage{subfigure}
\def \be {\begin{equation}}
\def \ee {\end{equation}}
\def \bea {\begin{eqnarray}}
\def \eea {\end{eqnarray}}
\numberwithin{equation}{section}
\title{\textbf{Shock Formation in Lovelock Theories}}

\author{Harvey S. Reall$^a$, Norihiro Tanahashi$^{b,a}$, Benson Way$^a$}
\affil{ \textit{$^a$DAMTP, Centre for Mathematical Sciences, University of Cambridge, Wilberforce Road, Cambridge CB3 0WA, UK}}
\affil{ 
\textit{$^b$Kavli Institute for the Physics and Mathematics of the Universe,}\\
\textit{Todai Institutes for Advanced Study, University of Tokyo (WPI),}\\
\textit{5-1-5 Kashiwanoha, Kashiwa, Chiba 277-8583, Japan}}

\date{}

\begin{document}
\maketitle

\begin{abstract}
We argue that Lovelock theories of gravity suffer from shock formation, unlike General Relativity. We consider the propagation of (i) a discontinuity in curvature, and (ii) weak, high frequency, gravitational waves. Such disturbances propagate along characteristic hypersurfaces of a ``background" spacetime and their amplitude is governed by a transport equation. In GR the transport equation is linear. In Lovelock theories, it is nonlinear and its solutions can blow up, corresponding to the formation of a shock. We show that this effect is absent in some simple cases e.g. a flat background spacetime, and demonstrate its presence for a plane wave background. We comment on weak cosmic censorship, the evolution of shocks, and the nonlinear stability of Minkowski spacetime, in Lovelock theories.
\end{abstract}

\onehalfspacing
%===MAIN=================================================================
\section{Introduction}

Lovelock theories of gravity \cite{lovelock} are natural alternatives to General Relativity (GR) in more than four spacetime dimensions.  These are higher curvature theories of gravity where the equations of motion remain second order in derivatives.  A well-known feature of Lovelock theories is that gravitational signals can propagate faster or slower than light \cite{aragone,CB}.  It is natural to ask whether or not the back of a wavepacket can catch up with the front and form a shock.  In this paper we will argue that Lovelock theories do suffer from such shock formation, unlike GR.

\begin{figure}[htbp]
\centering
\includegraphics[width=12cm]{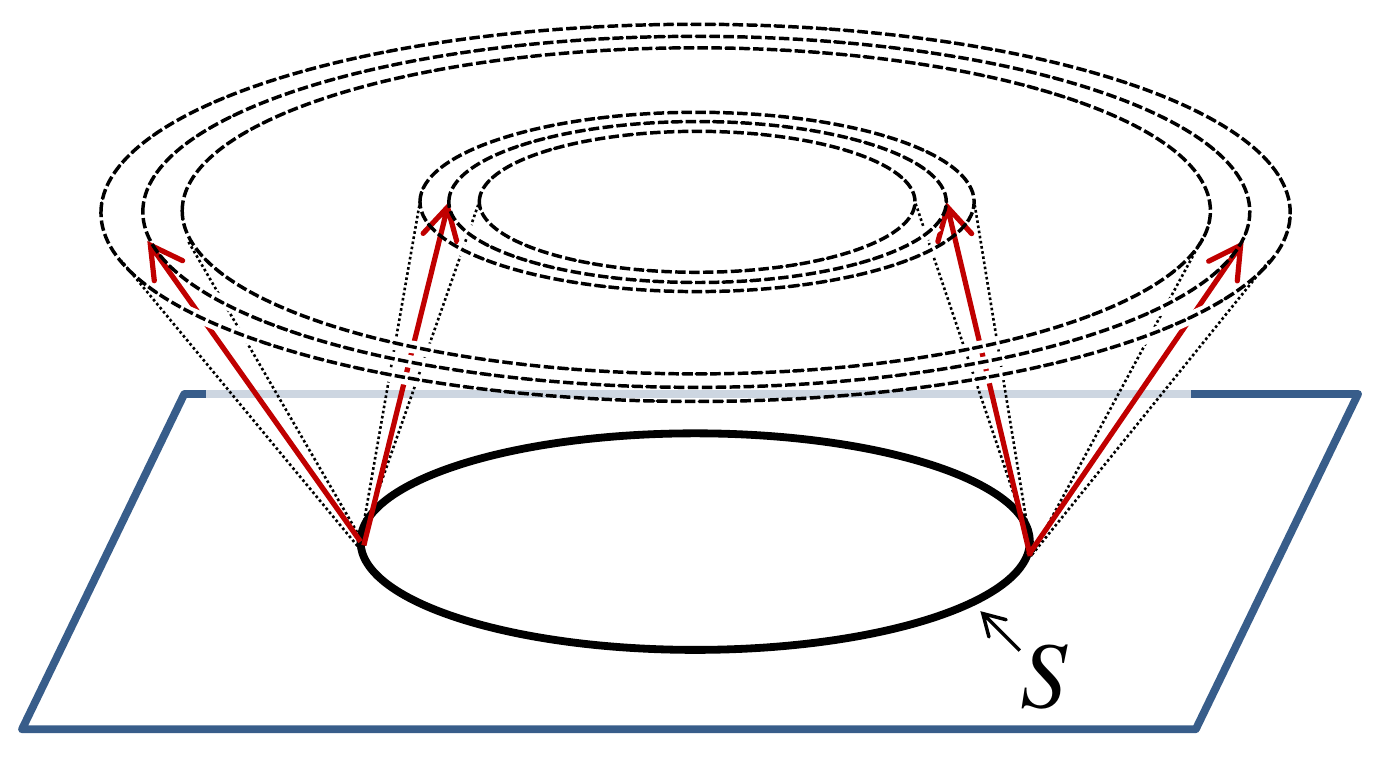}
\caption{
Characteristic hypersurfaces emanating from $(d-2)$-dimensional surface $S$.}
\label{Fig1}
\end{figure}

Causality of a physical theory is determined by its characteristic hypersurfaces. For example, given initial data specified on a suitable hypersurface $\Sigma$, the region of spacetime determined uniquely by the data inside a compact $(d-2)$-dimensional surface $S \subset \Sigma$ is bounded by a characteristic hypersurface emanating from $S$. In GR, a hypersurface is characteristic if, and only if, it is null. Characteristic surfaces in Lovelock theories have been discussed in Refs. \cite{aragone,CB,izumi,us}. Such surfaces are generically non-null so gravitational signals can travel faster (or slower) than light.\footnote{
For asymptotically anti-de Sitter or asymptotically flat boundary conditions, it has been argued that ``asymptotic causality" is violated in Einstein-Gauss-Bonnet theory (a Lovelock theory) \cite{brigante,camanho}. This means that a signal from the boundary can propagate through the bulk and return to the boundary faster than any signal that remains in the asymptotic region. This excludes the existence of a consistent dual CFT. However, it is not obvious that it implies pathological behaviour for the initial value problem in the bulk, see e.g. Ref. \cite{geroch}.} Furthermore, different polarizations of the graviton are associated to different characteristic surfaces i.e. they propagate with different speeds. For example, in Ref. \cite{us} we demonstrated that, for a certain class of spacetimes (Ricci flat with Weyl tensor of type N), generically there is a distinct ``ingoing'' and ``outgoing'' characteristic hypersurface emanating from $S$ for each polarization of the graviton. This is sketched in Fig. \ref{Fig1}. We expect this to be typical of the behaviour in a generic background, i.e., Lovelock theories are ``multirefringent''.\footnote{Ref. \cite{us} also showed that, when the curvature is comparable to the scale set by the Lovelock coupling constants, the theory can fail to be hyperbolic, in which case the initial value problem is ill-posed.}

Since the ``speed'' of a particular graviton polarization can vary in spacetime, one could construct initial data corresponding to a gravitational wavepacket in which the back of the wavepacket is moving faster than the front. This might result in wave steepening and shock formation. The canonical example of such behaviour is Burgers' equation
\be
\label{burgers}
 u_t + u\, u_x = 0\;,
\ee
where subscripts denote partial derivatives. For this equation, $u(t,x)$ is constant along characteristics which are straight lines with velocity $dx/dt = u$. If initial data is such that $u_x(t,x_0) < 0$ at some point $x_0$, then after a finite time, the characteristic emanating from $x=x_0$ will intersect its neighbours. When this happens, there is a blow-up of $u_x$. This is interpreted as shock formation.

A well-known method for investigating this phenomenon for PDEs of order $k$ is to consider the evolution of initial data with a discontinuity in the $k$th derivatives of the fields. As we will explain, such a discontinuity must propagate along a characteristic surface in spacetime. Characteristic surfaces are ruled by bicharacteristic curves (null geodesics in GR). The amplitude of a discontinuity is governed by a transport equation (an ODE), along a bicharacteristic curve. By solving this ODE, one can determine whether the amplitude can diverge at some finite time. 

A particularly interesting case is when the solution on one side of the discontinuity is known explicitly. We refer to this known solution as the ``background" solution. The discontinuity propagates along an ``outgoing'' characteristic hypersurface of the background solution, corresponding to a wavefront ``invading'' the region described by the background solution. In this case, the discontinuity transport equation depends only on the background solution. Hence, one can determine whether or not blow-up occurs without having to determine the solution on the other side of the discontinuity.

A closely related approach is to consider a nonlinear generalization of geometric optics \cite{CB3,hunter,anile,CBbook}. In this approach, one considers weak (small amplitude) high frequency waves propagating in a background solution. The surfaces of constant phase are characteristic hypersurfaces of the background solution. The waves are transported along the bicharacteristic curves within these surfaces according to a certain ODE that depends only on the background solution. By solving this equation one can determine whether or not blow-up of the waves occurs. 

The transport equations for the discontinuity and for weak, high frequency waves typically contain a nonlinear term.  The coefficient of this nonlinear term is the same for both equations.  For some physical theories, this nonlinear term vanishes for any background, i.e., the transport equations are always linear. Such theories are referred to as ``exceptional'' or ``linearly degenerate''.  If the ODEs are generically nonlinear then the theory is ``genuinely nonlinear''. 

GR is an exceptional theory. For example, the transport equation for a discontinuity in second derivatives of the metric, i.e., a curvature discontinuity is \cite{lich}
\be
 V^e \nabla_e [R_{abcd}] + \frac{1}{2} \theta [R_{abcd}] =0\;,
\ee
where $V^a$ is tangent to the affinely parameterized null geodesic generators of the null hypersurface along which the discontinuity propagates, $\theta = \nabla_a V^a$ is the expansion of these generators, and $[R_{abcd}]$ is the curvature discontinuity. Yang-Mills theory is also exceptional, and so is Born-Infeld theory \cite{boillat1970}. In contrast, a relativistic perfect fluid is genuinely nonlinear except for certain special equations of state such as a stiff fluid ($p=\rho$) \cite{CB3,boillat3}. 

In the exceptional case, a solution of the transport equations can diverge only if bicharacteristic curves intersect, i.e., at a caustic. If one arranges a discontinuity to travel along a characteristic surface which does not form a caustic, then the discontinuity will not diverge. For example, in GR, if we takes the ``background'' solution to be an asymptotically flat spacetime, and consider a caustic-free outgoing characteristic (i.e. null) hypersurface then the amplitude of a curvature discontinuity propagating along this surface will decrease as the discontinuity moves outwards. Weak high frequency waves behave similarly. 

%In the case of asymptotic waves this divergence just represents a breakdown of the approximation, analogous to the breakdown of geometric optics at a caustic. Similarly, in the case of a solution with a discontinuity, if one considers a sequence of smooth solutions that converges to the discontinuous solution then these smooth solutions typically would not blow up - the blow up occurs only in the limit.

In a genuinely nonlinear theory, for which the transport equations are nonlinear, a solution of these ODEs may diverge even when no caustics are present. This corresponds to shock formation. In the case of a discontinuity, the divergence occurs because faster-moving outgoing characteristic surfaces in the region ``behind'' the discontinuity eventually intersect (``catch up with'') the characteristic surface along which the discontinuity propagates. 

In an exceptional theory, since the transport equations are linear, the initial size of the discontinuity (or amplitude of the high frequency waves) does not affect its subsequent evolution beyond setting an overall scale. Blow up occurs only if a caustic forms. However, in a genuinely nonlinear theory, the nonlinear nature of these ODEs implies that one can ensure that the discontinuity or high frequency wave blows up simply by taking it to be large enough initially (with appropriate sign), irrespective of whether or not the characteristic surface along which it propagates contains a caustic. This corresponds to the formation of a shock from ``large'' initial data. 

One might wonder whether such divergences are an artifact of assuming a discontinuity in the initial data, or of making the high frequency approximation. But, in a genuinely nonlinear theory, similar behaviour can occur for exact solutions arising from {\it smooth} initial data. For example, for a large class of (systems of) genuinely nonlinear first order PDEs, it has been proved that there exist exact plane wave solutions, arising from smooth initial data of arbitrarily small amplitude, which blow up in finite time \cite{majda,john}. A detailed study of shock formation in fluids is given in Ref. \cite{christodouloufluids}, which proved that blow-up can occur for smooth, small amplitude, initial data for a relativistic perfect fluid in $3+1$ dimensions (extending earlier work of Ref. \cite{sideris} for a non-relativistic compressible perfect fluid). This blow-up is  associated to a divergence in the ``density'' of outgoing characteristic hypersurfaces, i.e. such surfaces ``catch up with each other'' as described heuristically above. 

In this paper we will investigate whether Lovelock theories suffer from shock formation. This amounts to asking whether such theories are exceptional or genuinely nonlinear. This question has been considered before, with conflicting results. Refs. \cite{Tomimatsu:1986wb,Tomimatsu:1987xy} considered a particular (toroidal) reduction of Lovelock theory to 1+1 dimensions and concluded that the resulting theory {\it is} exceptional. Ref. \cite{CB2} considered weak high frequency waves in Einstein-Gauss-Bonnet theory (a Lovelock theory) in a limit in which the Gauss-Bonnet coupling scales in inverse proportion to the frequency of the waves. The result was a {\it linear} ODE governing transport of the waves. Nevertheless, it was stated that the form of this ODE implies that the theory is not exceptional. As we will discuss below, this claim refers to a notion of exceptionality that is different (although related) to the one discussed above, and is not related to shock formation. 

We will show that Lovelock theories are genuinely nonlinear. We do this using the two methods described above. First, we consider solutions with a discontinuity in second derivatives of the metric, i.e., a discontinuity in curvature.  Second, we consider weak, high frequency, gravitational waves, but without the assumption of Ref. \cite{CB2} that the coupling constants scale with frequency. In both cases, we find a nonlinear transport equation. 

Since Lovelock theories are genuinely nonlinear, shock formation is generic.  Nevertheless, in a number of simple cases with special symmetries, we find that the nonlinear term in the transport equations vanish, and hence no shock formation occurs.  These include: (i) any characteristic hypersurface in a flat background (ii) spherically symmetric characteristic hypersurfaces in static, spherically symmetric backgrounds and (iii) Killing horizons (which are characteristic hypersurfaces in Lovelock theories \cite{izumi,us}).  

To obtain a tractable example of shock formation, we take our background spacetime to be a homogeneous plane wave. In this case, we find a characteristic hypersurface for which the nonlinear term is non-zero and there exist solutions of the transport equations which form shocks, starting from an initial disturbance of arbitrarily small amplitude. In this case, the hypersurface has a caustic which leads to focusing of the curvature discontinuity (or weak high frequency waves) and a shock forms before the caustic. 
 
 It is natural to ask whether our results have any implications for cosmic censorship, or for the stability of Minkowski spacetime, in Lovelock theories. We will discuss these issues at the end of this paper. This paper concerns the {\it formation} of shocks in Lovelock theories. It is an interesting question whether there is any sense in which a solution can be extended beyond shock formation. In particular, can one develop a theory of the {\it evolution} of shocks? This will also be discussed at the end of this paper. 
 
This paper is organised as follows.  In the following section, we discuss in general terms characteristic hypersurfaces and the transport of discontinuities and high frequency waves.  In section 3, we focus on Lovelock theories and demonstrate that the transport equations are generically nonlinear.  We then give a number of examples and finish with a discussion.  
 
 \section{Transport equations in 2nd order theories}
 
 \subsection{Characteristic hypersurfaces and bicharacteristic curves}
 
 In this section we will review the definitions of characteristic hypersurfaces and bicharacteristic curves \cite{courant}. Consider a field theory in $d$ spacetime dimensions, in which the unknown fields form a column vector $g_I$, $I = 1, \ldots, N$, with equation of motion
\be
 E_I \left( g,\partial g, \partial^2 g \right)=0
\;.\ee
(In a Lovelock theory $g_I$ will stand for the dynamical components of the metric.)  The theory is quasilinear if $E_I$ is linear in $\partial^2 g_J$. We will not assume this. However, as we will show in section 3, in any coordinate chart $x^\mu$, the Lovelock equations of motion depend linearly on $\partial_0^2 g_{\mu\nu}$ \cite{aragone}. So we will assume that $E_I$ has this property. Hence in any chart the equation of motion takes the form
\be
\label{eqofmotion}
A_{IJ} \partial_0^2 g_J + \cdots = 0
\;,\ee
where the ellipsis denotes terms involving fewer than 2 derivatives with respect to $x^0$ and $A_{IJ}$ does not depend on $\partial_0^2 g_J$. 

Now consider a hypersurface $\Sigma$ and introduce adapted coordinates $(x^0,x^i)$ so that $\Sigma$ has equation $x^0=0$. Assume that $g_I$ and $\partial_\mu g_I$ are known on $\Sigma$. By acting with $\partial_i$ we then also know $\partial_i \partial_\mu g_I$ on $\Sigma$. The only second derivatives that we don't know are $\partial_0^2 g_I$. These are uniquely determined by the equation of motion (\ref{eqofmotion}) if, and only if, the matrix $A_{IJ}$ is invertible. If this is the case then $\Sigma$ is said to be non-characteristic. If the matrix is {\it not} invertible anywhere on $\Sigma$ then $\Sigma$ is characteristic:
\be
 \det A = 0 \qquad \Leftrightarrow\qquad  \Sigma \; {\rm characteristic}\;.
\ee
To write things covariantly we define the {\it principal symbol} of the equation
\be
 P(x,\xi)_{IJ} = \frac{\partial E_I}{\partial (\partial_\mu \partial_\nu g_J)} \xi_\mu \xi_\nu
\ee
for an arbitrary covector $\xi_\mu$. We then have $A_{IJ} = P(x,dx^0)_{IJ}$. The {\it characteristic polynomial} is
\be
 Q(x,\xi) = \det P(x,\xi)
\;.\ee
$Q$ is a homogeneous polynomial in $\xi$. A hypersurface $\phi(x)={\rm constant}$ is characteristic iff $Q(x,d\phi)=0$ everywhere on the surface. This is a first order PDE for $\phi$. The theory of first order PDEs implies that such surfaces are generated by {\it bicharacteristic curves} $(x^\mu(s),\xi_\nu(s))$ defined by \cite{courant}
\be
\label{bicharacteristic}
 \dot{x}^\mu = \frac{\partial Q}{\partial \xi_\mu}\;, \qquad \dot{\xi}_\mu = -\frac{\partial Q}{\partial x^\mu}\;,
 \ee
with the initial values of $\xi_\mu$ chosen so that $Q=0$ (this is preserved along the curves). In GR, a hypersurface is characteristic if, and only if, it is null, and bicharacteristic curves are null geodesics. 

\subsection{Propagation of discontinuities}

\label{discont}

Consider a solution which is smooth everywhere except across a hypersurface $\Sigma$ on which the solution is $C^1$ but $\partial^2 g_I$ is discontinuous. In this case, the equations of motion cannot uniquely determine $\partial^2 g_I$ on $\Sigma$. Hence $\Sigma$ must be a characteristic surface. In adapted coordinates, the discontinuous components of $\partial^2 g_I$ are $\partial_0^2 g_I$ since, as discussed above, the components $\partial_i \partial_\mu g_I$ are determined uniquely. Notice that $A_{IJ}$ is continuous because it does not depend on $\partial_0^2 g_I$. Therefore, $\det {\bf A}$ is also continuous so the hypersurface is characteristic w.r.t. the solution on both sides of $\Sigma$.\footnote{By taking derivatives of the equation of motion one can see that discontinuities in $\partial^k g_I$, $k \ge 3$ also propagate along characteristic hypersurfaces, i.e., if a solution is smooth on either side of $\Sigma$ and $C^{k-1}$ on $\Sigma$ with a discontinuity in $\partial^k g_I$ on $\Sigma$ then $\Sigma$ must be characteristic.}

Taking the discontinuity in the equation of motion across $\Sigma$ gives
\be
 A_{IJ} [\partial_0^2 g_J] = 0\;,
\ee
where square brackets denote the discontinuity. Hence $[\partial_0^2 g_J]$ is an eigenvector of $A_{IJ}$ with eigenvalue $0$, i.e., it is an element of the kernel of ${\bf A}$. Write this element as $r_J$ and let $\xi =dx^0$ be the normal to hypersurface. Written covariantly we have
\be
\label{discg}
 [\partial_\mu \partial_\nu g_I] = \xi_\mu \xi_\nu r_I\;,
\ee
where $r_I$ is an element of the kernel of $P(x,\xi)$. 

Now consider initial data specified on a non-characteristic hypersurface, such that the data has a discontinuity in $\partial^2 g_I$ across a $(d-2)$ dimensional surface $S$ within this hypersurface. In the resulting solution, any discontinuity must propagate along a characteristic hypersurface. In a general second order hyperbolic theory with $N$ degrees of freedom, there will be $2N$ characteristic surfaces emanating from $S$: an ``outgoing'' and an ``ingoing'' characteristic surface for each degree of freedom (see Fig.\ref{Fig1}). 
%{\bf FIGURE} 
In GR, the outgoing surfaces are all coincident and the ingoing surfaces are all coincident because all gravitational degrees of freedom propagate at the speed of light (surfaces are characteristic if and only if they are null). However, as discussed in the Introduction, this is not true in Lovelock theories. 

In general, the discontinuity in the initial data at $S$ will lead to discontinuities propagating along each of the characteristic surfaces through $S$. A particularly interesting case is when $S$ divides the initial data surface into two regions and the initial data on one side of $S$ corresponds to a known explicit solution, which we will call the ``background'' solution. We will refer to this side of $S$ as the ``outside'' and the other side as the ``inside'' with a corresponding division of the characteristic surfaces into ``outgoing'' and ``ingoing''. Everywhere outside the outermost outgoing characteristic surface, the solution will coincide with the background solution. Inside this characteristic surface, the solution will depend on the initial data inside $S$. Hence this characteristic surface is a wavefront ``invading'' the region of spacetime described by the background solution. We will focus on the amplitude of the discontinuity propagating along this outermost outgoing characteristic surface since, as we will show, it satisfies a transport equation that can be determined from the form of the background solution.

A useful reference for the propagation of discontinuities is Ref. \cite{anile}. What follows is an application of the methods described there to the class of theories described above. 
%The theory is originally due to Lax, Choquet-Bruhat, Boillat,...

%Let's discuss this is a little more detail. Denote the characteristic surfaces through $S$ as $\Sigma^{I\pm}$ ($I=1, \ldots, N$) with normals $\xi_\mu^{(I\pm)}$ where $\pm$ refers to outgoing and ingoing respectively. Then, since the principal symbol $P(x,\xi^{(I\pm)})=0$ is degenerate, there exists a vector $r^{(I\pm)}$ such that $P(x,\xi^{(I\pm)}) \cdot r^{(I\pm)} = 0$. We now decompose our initial discontinuity on $S$ into a sum of the discontinuities associated with each characteristic surface:
%\be
 %[\partial_\mu \partial_\nu g_I]_S = \sum_J \Pi^{(J+)} \xi_\mu^{(J+)} \xi_\nu^{(J+)} r^{(J+)}_I + \sum_J \Pi^{(J-)}\xi_\mu^{(J-)} \xi_\nu^{(J-)} r^{(J-)}_I
%\ee
%for some coefficients $\Pi^{(J\pm)}$. These coefficients represent the initial amplitude of the discontinuity propagating along each characteristic surface. 

Introduce coordinates $(x^0,x^i)$ adapted to the outermost outgoing characteristic hypersurface $\Sigma$ with $x^0<0$ corresponding to the ``background'' where the solution is known explicitly. We assume the equation of motion takes the form
\be
\label{evolution}
A_{IJ}({\bf g}_{ij}, {\bf g}_0,{\bf g}_i,{\bf g},x) (g_J)_{00} + b_I({\bf g}_{0i}, {\bf g}_{ij}, {\bf g}_0, {\bf g}_i, {\bf g},x) = 0\;.
\ee
Here subscripts $0,i$  denote partial derivatives w.r.t $x^0,x^i$ and ${\bf g}(x)$ is a vector with components $g_I$. Note that we are now assuming that $A_{IJ}$ does not depend on ${\bf g}_{0i}$, which was not assumed above but is true for Lovelock theories, as we will show in section 3. 

The characteristic condition for our surface $\Sigma$ with equation $x^0=0$ is $\det {\bf A} = 0$, which implies that ${\bf A}$ admits left and right eigenvectors $l_I$ and $r_J$ with eigenvalue $0$:
\be
\label{evolution2}
 l_I A_{IJ} = A_{IJ} r_J = 0 \qquad (x^0=0)\;.
\ee
We will assume that the eigenvalue $0$ is non-degenerate so that $l_I$, $r_I$ are unique up to scaling. In Lovelock theories, we showed that this is true for a generic Ricci flat type N spacetime \cite{us} and we believe it to be true generically. 

We now allow for a discontinuity in second derivatives across $\Sigma$. As explained above, the discontinuous components are $(g_I)_{00}$ and these must be proportional to $r_I$:
\be
\label{Pidef}
 [(g_I)_{00}] = \Pi r_I
\ee
for some scalar $\Pi(x^i)$ defined on $\Sigma$. Here we assume that $r_I$ has been normalized in some way so that $\Pi$ gives a measure of the size of the discontinuity. 

To obtain an evolution equation for $\Pi$ we take a $x^0$-derivative of (\ref{evolution}), evaluate at $x^0=0$, and contract with $l_I$ to eliminate 3rd derivatives w.r.t. $x^0$. This gives
\be
\label{evolutiont}
l_{I} \left\{ (A_{IJ})_0 (g_J)_{00} + (b_I )_0 \right\} = 0 \qquad (x^0=0)\;.
\ee
Now we use the chain rule:
\be
 (A_{IJ})_0  = \frac{\partial A_{IJ}}{\partial (g_K)_{ij}} (g_K)_{0ij} + \frac{\partial A_{IJ}}{\partial (g_K)_{0}} (g_K)_{00} + \frac{\partial A_{IJ}}{\partial (g_K)_{i}} (g_K)_{0i} + \frac{\partial A_{IJ}}{\partial (g_K)} (g_K)_{0} + \frac{\partial A_{IJ}}{\partial x^0}
\ee
\be
 (b_I)_0 = \frac{\partial b_I}{\partial (g_J)_{0i}} (g_J)_{00i} + \frac{\partial b_I}{\partial (g_J)_{0}} (g_J)_{00}+\frac{\partial b_I}{\partial (g_J)_{ij}} (g_J)_{0ij}+\frac{\partial b_I}{\partial (g_J)_{i}} (g_J)_{0i}+\frac{\partial b_I}{\partial (g_J)} (g_J)_{0} + \frac{\partial b_{I}}{\partial x^0}\;,
\ee
where the final term of these equations arises from the explicit $x^0$ dependence of $A_{IJ}$ and $b_I$, if present. 
Substituting this into (\ref{evolutiont}) and taking the discontinuity gives
\be
\label{evolutiondisc}
 l_I \left\{  \frac{\partial b_I}{\partial (g_J)_{0i}} [(g_J)_{00i}] + \frac{\partial A_{I J}}{\partial (g_K)_0} [(g_K)_{00} (g_J)_{00}] + B_{I J} [( g_J)_{00} ] \right\}= 0\;,
\ee
where
\be
 B_{IJ} = \frac{\partial A_{IJ}}{\partial (g_K)_{ij}} (g_K)_{0ij} +  \frac{\partial A_{I J}}{\partial (g_K)_{i}} (g_K)_{0i} + \frac{\partial A_{I J}}{\partial (g_K)} (g_K)_{0}
+ \frac{\partial A_{IJ}}{\partial x^0}
+  \frac{\partial b_I}{\partial (g_J)_{0}} \;.
 \ee
Let $(g_I)_{00}^- = \lim_{x^0 \rightarrow 0^-} (g_I)_{00}$. Then we have
\be
  [(g_I)_{00} (g_J)_{00}]  = [(g_I)_{00} ][(g_J)_{00}] + [(g_I)_{00} ](g_J)^-_{00} + (g_I)^-_{00} [(g_J)_{00}] \;.
\ee
Using this, and (\ref{Pidef}), equation (\ref{evolutiondisc}) becomes
\be
\label{mastereq1}
 K^i \Pi_i + N \Pi^2 + M \Pi =0 \;,
\ee 
where
\be
\label{Kdef}
K^i =  l_I \frac{\partial b_I}{\partial (g_J)_{0i}} r_J\;,
\ee
\be
\label{Ndef}
 N= l_I  \frac{\partial A_{I J}}{\partial (g_K)_0} r_J r_K\;,
 \ee
 and
\bea
\label{Mdef}
M &=& l_I \left\{  \frac{\partial b_I}{\partial (g_J)_{0i}} (r_J)_i + \left(
  \frac{\partial A_{I J}}{\partial (g_K)_{ij}} (g_K)_{0ij} +  \frac{\partial A_{I J}}{\partial (g_K)_{i}} (g_K)_{0i} + \frac{\partial A_{I J}}{\partial (g_K)} (g_K)_0
+ \frac{\partial A_{IJ}}{\partial x^0}
+  \frac{\partial b_I}{\partial (g_J)_{0}} \right) r_J \right. \nonumber \\&+&\left. \frac{\partial A_{I J}}{\partial (g_K)_0} \left( (g_K)^-_{00} r_J + (g_J)^-_{00} r_K \right) \right\}\;.
\eea
Equation (\ref{mastereq1}) is an ODE along the integral curves of $K^i$, which lie within $\Sigma$ ($x^0=0$). Let $s$ be a parameter along such a curve, i.e.,
\be
\label{sdef}
 \frac{dx^i}{ds} = K^i(x^j)\;,
\ee
then (\ref{mastereq1}) becomes
\be
\label{mastereq}
 \dot{\Pi} + N \Pi^2 + M \Pi =0 \;,
\ee 
where a dot denotes a derivative w.r.t. $s$. Note that $N$ and $M$ can be determined by the limiting behaviour of the background solution as $x^0 \rightarrow 0^-$. Hence the transport equation (\ref{mastereq}) for the discontinuity depends only on the form of this background solution. 

We will now show that the curves $x^i(s)$ are the bicharacteristic curves which generate $\Sigma$.  The principal symbol of (\ref{evolution}) is 
\be
 P(x,\xi)_{IJ} = A_{IJ} (\xi_0)^2 +2 \frac{\partial b_I}{\partial (g_J)_{0i}} \xi_0 \xi_i + \left( \frac{\partial A_{IK}}{\partial (g_J)_{ij} } (g_K)_{00} + \frac{\partial b_I}{\partial (g_J)_{ij}} \right) \xi_i \xi_j\;.
\ee
Bicharacteristic curves $(x^\mu(s),\xi_\mu(s))$ are determined by (\ref{bicharacteristic}). For the bicharacteristic generators of $\Sigma$ we have $x^0=0$, $\xi_i = 0$, $\xi_0 \ne 0$. To evaluate the derivative of $Q$ in (\ref{bicharacteristic}) we use %{\bf the result of this argument needs checking!}
 \be
  \frac{\partial Q}{\partial \xi_\mu}  = ({\rm adj} P)_{IJ} \frac{\partial P_{JI}}{\partial \xi_\mu}\;,
 \ee
 where ${\rm adj} P$ is the adjugate matrix of $P$ (the transpose of the cofactor matrix). Hence evaluating at $x^0=0$ gives
 \be
  \dot{x}^i = 2 {\rm adj} (\xi_0^2 A)_{IJ} \xi_0  \frac{\partial b_J}{\partial (g_I)_{0i}}\;.
 \ee
At $x^0=0$ we know that $A$ has left and right eigenvectors $l_I$ and $r_J$ with eigenvalue $0$. This implies that $({\rm adj} A)_{IJ} \propto l_I r_J$ and hence
\be
\label{bicharacteristicx}
 \dot{x}^i \propto l_I  \frac{\partial b_J}{\partial (g_I)_{0i}} r_J = K^i\;.
\ee
We can make this expression an equality by an appropriate choice of the parameter $s$ along the bicharacteristic curves. Hence solutions of (\ref{sdef}) (with $x^0=0$) are indeed the bicharacteristic curves which generate $\Sigma$.

Equation (\ref{mastereq}) is our transport equation, an ODE governing the propagation of the discontinuity along the bicharacteristic curves of $\Sigma$. In general, $N\neq0$ so this equation is nonlinear. However, some theories have the special property that $N$ vanishes for any background solution. Such theories are referred to as ``exceptional'' or ``linearly degenerate''.\footnote{An equivalent definition states that the derivatives of $Q(x,\xi)$ w.r.t. $x^\mu$ should be continuous across a characteristic surface along which a discontinuity propagates \cite{boillat2}.} If $N$ is generically non-zero then the theory is called ``genuinely nonlinear''.

Along any bicharacteristic curve, equation (\ref{mastereq}) has the general solution \cite{anile}
\be
\label{Pisoln}
 \Pi(s) = \Pi(0) e^{-\Phi(s)} \left( 1 + \Pi(0) \int_0^s N(s') e^{-\Phi(s') } ds' \right)^{-1}  \;,
\ee 
where $\Phi$ is defined by
\be
\label{Phidef}
 \Phi(s) = \int_0^s M(s') ds'\;.
\ee
Now we can ask whether $\Pi(s)$ can blow up at finite $s$. In an exceptional theory ($N=0$), the only way that this can happen is if  $e^{-\Phi(s)}$ blows up at finite $s$. Since $\Phi$ is determined entirely by the background solution, this can happen only if the background solution is not smooth, or the characteristic surface $\Sigma$ is not smooth. Assuming that the background is smooth and that $S$ is smooth, the only way that $\Sigma$ can fail to be smooth is if nearby bicharacteristic curves within $\Sigma$ intersect, i.e., $\Sigma$ forms a caustic. If $S$ is chosen so that $\Sigma$ is free of caustics then $\Pi(s)$ will not blow up. Note that this statement is independent of the initial amplitude $\Pi(0)$ because (\ref{mastereq}) is linear in an exceptional theory.

\begin{figure}[htbp]
\centering
\includegraphics[width=12cm]{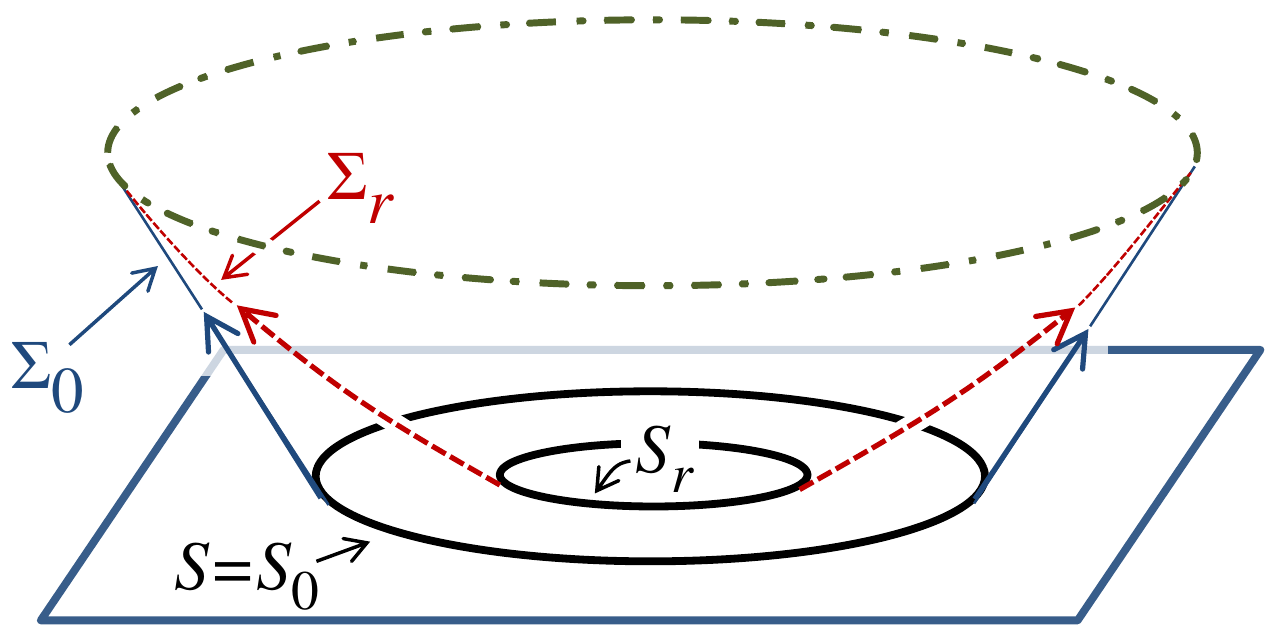}
\caption{
Foliation by $S_r$ and the outermost outgoing characteristic surfaces $\Sigma_r$.
If $\Sigma_r$ intersects $\Sigma_0$ for infinitesimal $r$,
a shock forms at the intersection (green dot-dashed curve).
}
\label{Fig2}
\end{figure}

In a genuinely nonlinear theory,  $\Pi(s)$ will diverge if
\be
\label{shockcondition}
1 + \Pi(0) \int_0^s N(s') e^{-\Phi(s') } ds' \rightarrow 0 \;.
\ee 
This is a nonlinear effect. It corresponds to the formation of a shock. Shock formation can be understood heuristically as follows. On the initial data surface, consider foliating the interior of $S$ with surfaces diffeomorphic to $S$. Denote this foliation by $S_r$, $r \ge 0$ where $S_0 =S$. From each $S_r$, let $\Sigma_r$ denote the outermost outgoing characteristic surface, so $\Sigma_0 = \Sigma$. A shock forms when, for infinitesimal $r$, $\Sigma_r$ intersects $\Sigma_0$. See Fig. \ref{Fig2}. 
The shock forms because the disturbance behind the wavefront travels faster than the front itself. This heuristic picture is supported more rigorously by studies of genuinely nonlinear first order systems in $1+1$ dimensions (the simplest being the equation (\ref{burgers})) \cite{majda,john}, and by Christodoulou's analysis of shock formation in relativistic perfect fluids in $3+1$ dimensions \cite{christodouloufluids}. In this work, he proved that shock formation corresponds to the divergence of a certain ``density'' of outgoing characteristic surfaces. 

In an exceptional theory, the initial amplitude $\Pi(0)$ does not affect the subsequent evolution of the discontinuity, beyond setting a scale. In contrast, in a genuinely nonlinear theory, shock formation can be guaranteed simply by taking $\Pi(0)$ large enough, with appropriate sign. To see this, assume that $N \ne 0$ at some point on $S$ and consider the bicharacteristic curve in $\Sigma$ that passes through this point. Along this curve, $N(s) e^{-\Phi(s)}$ has a definite sign for small $s$. Hence the magnitude of the integral in (\ref{shockcondition}) is monotonically increasing for small $s$. Therefore by taking $\Pi(0)$ to be large enough, with opposite sign to $N$, we can ensure that (\ref{shockcondition}) occurs for small $s$. 

Above we assumed a discontinuity in second derivative across $\Sigma$. We could instead consider continuous second derivatives and a discontinuity in {\it third} derivatives. A transport equation for such a discontinuity is obtained by differentiating (\ref{evolution}) {\it twice} w.r.t. $x^0$, contracting with $l_I$ and setting $x^0=0$. In contrast to the above equation, this ODE is {\it linear}. This seems to be the reason why Refs. \cite{Tomimatsu:1986wb,Tomimatsu:1987xy} found that Lovelock theories are exceptional when toroidally reduced to 1+1 dimensions. These papers reformulated the equations as a first order system, which involved introducing second derivatives of the metric as fields. In a first order system, one considers a discontinuity in first derivatives of the fields. This corresponds to a discontinuity in third derivatives of the metric. As just explained, a discontinuity in third derivatives propagates according to a linear equation, even in a genuinely nonlinear theory.

\subsection{Weak high frequency waves}

\label{waves}

Here we consider high frequency, small amplitude, waves propagating in a ``background'' field, and perform an expansion in inverse powers of the frequency. This is very similar to the WKB approximation, or geometric optics, except that it is nonlinear. More detailed discussion of this approach can be found in Refs. \cite{CB3,hunter,anile,CBbook}. 

We will continue to work with the theory defined by (\ref{evolution}) and specialize to Lovelock theories in the next section. Introduce coordinates $x^\mu=(x^0,x^i)$ with the idea that surfaces of constant $x^0$ are, to leading order, surfaces of constant phase for the waves. We then make the {\it Ansatz}
\be
 g_I(x) = \bar{g}_I (x) + \omega^{-2} h_I (x,\eta) + \omega^{-3} \kappa_I (x,\eta) + \mathcal O(\omega^{-4})\;,
\ee 
where $\eta = \omega x^0$, subscripts denote partial derivatives, and we are interested in large $\omega$. The idea is that the solution depends on the ``slow'' coordinates $x^\mu$ associated with the background field $\bar{g}_I$, and also the ``fast'' coordinate $\eta$ associated with the oscillation of the waves. In particular, note that the dependence on $\eta$ determines the dependence on $\omega$. This is very similar to geometric optics except that we do not assume a factorized form for $h_I$. 

We assume that $h_I$, $k_I$ and their derivatives w.r.t. $\eta$ are uniformly bounded functions of $\eta$. Substituting into (\ref{evolution}) and expanding in $\omega^{-1}$ gives, at order $\omega^0$:
\be 
  {\bf A}[\bar{\bf g}] \cdot \left(  \bar{\bf g}_{00} + {\bf h}'' \right) + {\bf b}[\bar{\bf g}] =0\;,
\ee
where a prime denotes a derivative w.r.t. $\eta$ and subscripts $0$ or $i$ denote derivative w.r.t. $x^0$ or $x^i$ at fixed $\eta$. Now average $\eta$ over the interval $[0,T]$, i.e., act with $T^{-1} \int_0^T d\eta$. Let $T \rightarrow \infty$. Our boundedness assumption implies that the ${\bf h}''$ term drops out, giving
\be
 {\bf A}[\bar{\bf g}] \cdot \bar{\bf g}_{00}  + {\bf b}[\bar{\bf g}] =0\;,
\ee
i.e. the ``background'' $\bar{\bf g}$ must satisfy the equation of motion. Plugging back in above then gives
\be
 {\bf A}[\bar{\bf g}] \cdot {\bf h}''=0\;.
\ee
Hence if ${\bf h}'' \ne 0$ then ${\bf A}[\bar{\bf g}]$ must be degenerate:
\be
 \det {\bf A}[\bar{\bf g}] = 0\;,
\ee
which means that surfaces of constant $x^0$, i.e., surfaces of constant phase, are characteristic. Furthermore we must have
\be
 h_I''(x,\eta) =\Omega''(x,\eta) r_I(x)
\ee
for some function $\Omega''$, where $r_I$ is a right eigenvector of ${\bf A}[\bar{\bf g}]$ with eigenvalue $0$ (which we assume to be non-degenerate, as above). Integrating w.r.t. $\eta$ we obtain
\be
\label{hsol}
 h_I(x,\eta) = \Omega(x,\eta) r_I(x)\;,
\ee
where our boundedness assumption implies the absence of a term linear in $\eta$, and we have assumed, for simplicity, the vanishing of an $\eta$-independent part (this could be included but makes some of the equations longer). 
 
 Now consider the ${\cal O}(\omega^{-1})$ term in the equation of motion. This gives
\be
\frac{\partial A[\bar{\bf g}]_{IJ}}{\partial (\bar g_K)_0}h'_K\Bigl((\bar 
g_J)_{00}+h''_J\Bigr)+A[\bar{\bf g}]_{IJ}\Bigl(2(h_J')_0+\kappa_J''\Bigr)+\frac{\partial 
b[\bar{\bf g}]_I}{\partial (\bar g_J)_{0i}}(h'_J)_i
+\frac{\partial b[\bar{\bf g}]_I}{\partial (\bar g_J)_0}h'_J=0\;.
\ee
To eliminate dependence on $\kappa$, we now contract with $l_I$, a left eigenvector of ${\bf A}[\bar{\bf g}]$ with eigenvalue $0$, to obtain
\be
\label{OmegaPDE}
 K^i{\Omega}'_i + N \Omega' \Omega'' + \tilde{M} \Omega' = 0\;,
 \ee
where $K^i$ is defined by (\ref{Kdef}), $N$ is defined by (\ref{Ndef}) and 
\be
\label{tildeMdef}
 \tilde{M} =
 l_I\left(
\frac{A[\bar g]_{IJ}}{\partial (\bar g_K)_0}(\bar g_J)_{00}r_K
+ \frac{\partial b[\bar g]_I}{\partial (\bar g_J)_{0i}}(r_J)_i
+ \frac{\partial b[\bar g]_I}{\partial (\bar g_J)_0}r_J
\right)
\,.
\ee
Equation (\ref{OmegaPDE}) is a first order PDE for $\Omega'$. It constrains the dependence of $\Omega'$ on both $(x^0,x^i)$ {\it and} $\eta$. We solve this equation by the method of characteristics \cite{anile}. Consider curves $(x^\mu(s),\eta(s))$ defined by
\be
 \frac{dx^0}{ds} = 0\qquad \frac{dx^i}{ds} = K^i \qquad \frac{d\eta}{ds} = N\Omega'\;.
\ee
Equation (\ref{OmegaPDE}) reduces to the ODE
\be
\label{Omegapeq}
\frac{d\Omega'}{ds} + \tilde{M} \Omega'=0\;.
\ee
Note that $x^0$ is constant and $x^i(s)$ are simply the bicharacteristic curves within the surfaces of constant $x^0$. But we must also take account of the fact that $\eta$ evolves along these curves. 

\begin{figure}[htbp]
\centering
\includegraphics[width=8cm]{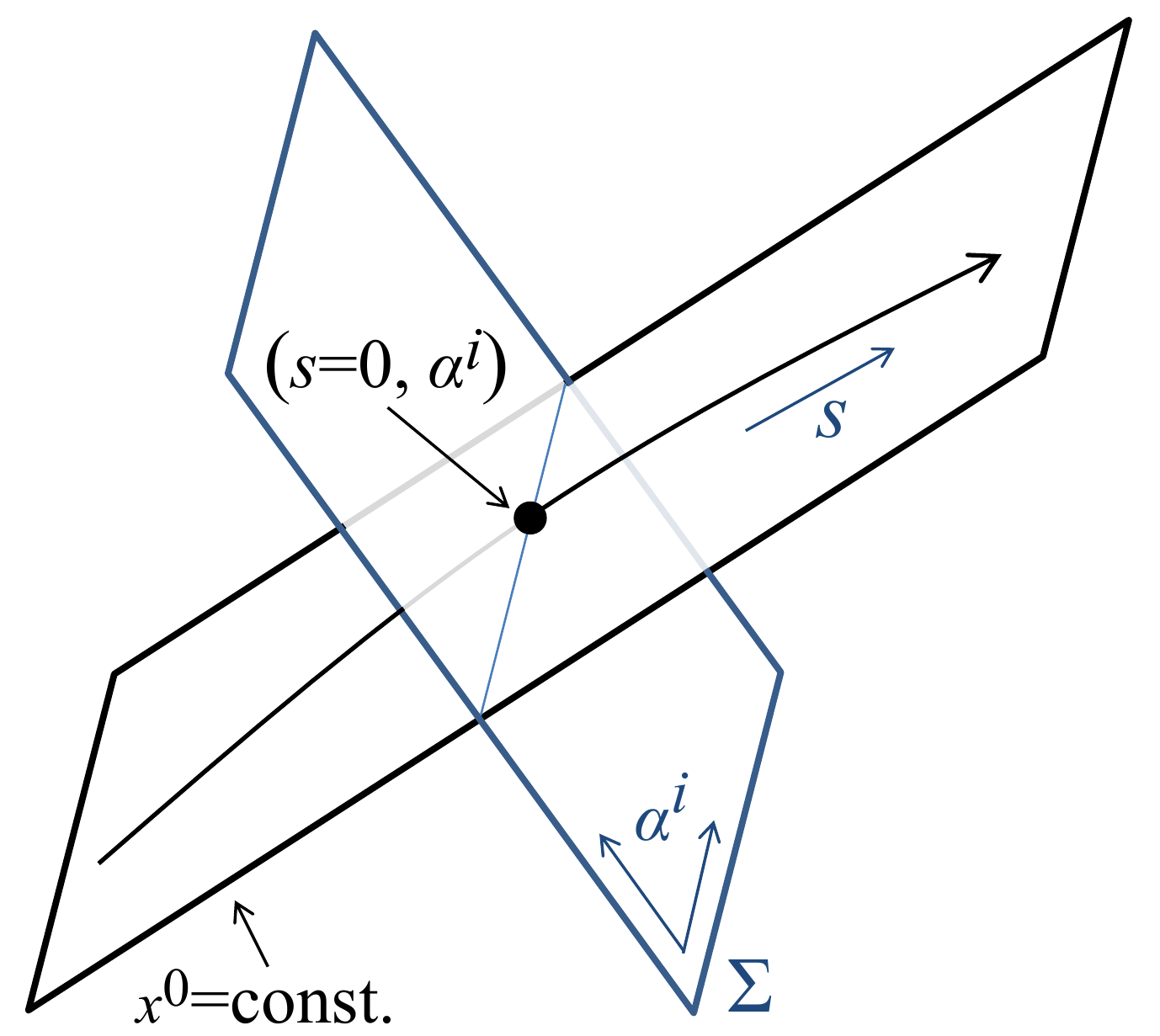}
\caption{
Parametrization of the bicharacteristic curves.
Parameterizing $\Sigma$, a surface transverse to the characteristic surfaces of constant $x^0$,
by $(s=0,\alpha^i)$ ($i=1,\ldots,d-1$), the bicharacteristic curves are given by
$x^0 = x^0(\alpha)$, $x^i=x^i(s,\alpha)$.
}
\label{Fig3}
\end{figure}

To solve this system of ODEs, pick a surface $\Sigma$ transverse to the (characteristic) surfaces of constant $x^0$.  Let $\alpha^i$ ($i=1,\ldots, d-1$) be coordinates on $\Sigma$ so that $\Sigma$ is given parametrically by $x^\mu=x^\mu(\alpha)$. Define the parameter along the bicharacteristic curves so that $s=0$ on $\Sigma$. Consider a bicharacteristic curve that intersects $\Sigma$ at the point with coordinates $\alpha^i$. Then this curve will be given by $x^0=x^0(\alpha)$, $x^i=x^i(s,\alpha)$ (since $x^0$ is constant along the curve). See Fig. \ref{Fig3}. As $\alpha^i$ varies, these bicharacteristic curves define a congruence in a region of spacetime. 

Let $\eta(0)=\beta$. Then, at last for small $s$, there is a unique solution to the above ODE specified by $(s,\alpha,\beta)$. 
The general solution to (\ref{Omegapeq}) is
\be
\label{Omegapsol}
 \Omega'(s,\alpha,\beta) = \Omega'(0,\alpha,\beta) e^{-\tilde{\Phi}(s,\alpha)}\;,
\ee
where (cf (\ref{Phidef}))
\be
 \tilde{\Phi}(s,\alpha) = \int_0^s \tilde M(x(s',\alpha)) ds' 
\ee
and $\Omega'(0,\alpha,\beta)$ is determined by initial conditions at $s=0$. We can now solve for $\eta$:
\be
\label{betadef}
 \eta(s,\alpha,\beta) = \beta + \Omega'(0,\alpha,\beta) I(s,\alpha)\;,
\ee
where
\be
 I(s,\alpha) = \int_0^s N(x(s',\alpha)) e^{-\tilde{\Phi}(s',\alpha)}  ds'\;.
\ee
We have solved (\ref{OmegaPDE}) and specified the solution parametrically in terms of $(s,\alpha,\beta)$. The final step is to perform a change of variables from $(s,\alpha,\beta)$ to $(x,\eta)$ so that we can express $\Omega'$ as a function of $(x,\eta)$. 

First consider the exceptional case $N \equiv 0$. Then we have $\beta=\eta$. In a region without caustics, the relation between $(s,\alpha)$ and $x^\mu$ is smooth and invertible so we can write
\be
 \Omega' = \Omega'(0,\alpha(x),\eta) e^{-\tilde{\Phi}(s(x),\alpha(x))} \qquad {\rm (exceptional)}\;,
\ee
where $\eta=\omega x^0$. Recall that $ \Omega'(0,\alpha(x),\eta) $ is determined by initial conditions. Since $\alpha(x)$ is constant along bicharacteristic curves, the first factor above does not change along these curves. 
In particular, consider a choice of initial data so that $\Omega'(0,\alpha,\beta)=W_1(\alpha) W_2(\beta)$, a function of position times a function of phase. Then the resulting solution also factorizes: $\Omega'(x,\eta) = {\cal A}(x) W_2(\eta)$ where ${\cal A} = W_1 e^{-\tilde{\Phi}}$. So the solution is given by a slowly varying (i.e. $\eta$-independent) amplitude ${\cal A}$ times a fixed phase factor $W_2$, just as in linear geometrical optics. Hence the dependence on $\eta$ does not change in time so there is no ``distortion of signals'' \cite{CB3}. The solution fails to be smooth only at a caustic, where derivatives of $\Omega'$ will blow up. This is also just as in geometrical optics.

Now consider the genuinely nonlinear case $N \ne 0$ and assume that there is no caustic in the region of interest. Then we can write $(s,\alpha)$ in terms of $x^\mu$ as above. But now we have to solve (\ref{betadef}) to determine $\beta$ as a function of $\eta$, $s$ and $\alpha$ and hence as a function of $\eta,x^\mu$. We can do this as long as $(\partial_\beta \eta)_{s,\alpha} \ne 0$. We have
\be
\label{deta}
( \partial_\beta \eta)_{s,\alpha} = 1 + \Omega''(0,\alpha,\beta) I(s,\alpha)
\ee
which is non-zero for small $s$ (because $I(0,\alpha)=0$). Hence, at least for small $s$ we can write
\be
 \Omega' = \Omega'(0,\alpha(x),\beta(s(x),\alpha(x),\eta)) e^{-\tilde{\Phi}(s(x),\alpha(x))}\;,
\ee
with $\eta=\omega x^0$. As above, consider the case for which the initial data factorizes into a function of position and a function of phase: $\Omega'(0,\alpha,\beta)=W_1(\alpha) W_2(\beta)$. The full solution takes the form $\Omega'(x,\eta) = {\cal A}(x)W_2(\beta(x,\eta))$. Since the second factor depends on $x$, the factorized form is {\it not} preserved in time evolution: ``signals are distorted'' in a genuinely nonlinear theory \cite{CB3}.  

For larger values of $s$, it might happen that the RHS of (\ref{deta}) vanishes at some finite value of $s$, beyond which we can no longer determine $\beta$ in terms of $\eta$, $s$ and $\alpha$. As we approach this value of $s$, $\partial_\eta \beta = 1/\partial_\beta \eta$ becomes large, i.e., $\beta$ is a rapidly varying function of $\eta$ at fixed $s,\alpha$. This implies that $\Omega'(x,\eta)$ develops a large gradient w.r.t. $\eta$ at fixed $x$: the profile of the waves ``become very steep''. To see this more precisely, consider\footnote{Note the exact correspondence with (\ref{Pisoln}). A discontinuity in second derivatives corresponds to a limit in which the weak high frequency waves approximation becomes exact.}
\be
\label{Omegasol}
 \Omega''(s,\alpha,\beta) = (\partial_\eta \Omega')_x= (\partial_\eta \Omega')_{s,\alpha}=\frac{(\partial_\beta \Omega')_{s,\alpha}}{(\partial_\beta \eta)_{s,\alpha}} = \frac{\Omega''(0,\alpha,\beta) e^{-\tilde{\Phi}(s,\alpha)}}{1 + \Omega''(0,\alpha,\beta) I(s,\alpha) }\;.
\ee
This expression blows up when (\ref{deta}) vanishes. This corresponds to shock formation. 

As long as $N \ne 0$ we can arrange that a shock forms by choosing the initial data appropriately. If $N \ne 0$ for $s=0$ and some $\alpha$ then $|I(s,\alpha)|$ is monotonically increasing for small positive $s$. Hence, by choosing the initial data $\Omega(0,\alpha,\beta)$ so that $\Omega''(0,\alpha,\beta)$ is large enough, with appropriate sign, we can arrange that (\ref{deta}) vanishes for small $s$, i.e., a shock forms at small $s$. There is no analogue of this for an exceptional theory. 

The distortion effect discussed above is the reason why Ref. \cite{CB2} claimed that Lovelock theories are genuinely nonlinear. As discussed in the Introduction, Ref. \cite{CB2} considered weak high frequency waves in Einstein-Gauss-Bonnet theory, taking the Gauss-Bonnet coupling to be of order $\omega^{-1}$. This results in a {\it linear} ODE governing the transport of such waves along bicharacteristics. This ODE includes a term linear in $\Omega''$. Such an equation can be solved as above \cite{anile}, with the result that $\eta = \beta + \Psi(s,\alpha)$ for some function $\Psi$. Since $\beta \ne \eta$, this still leads to ``distortion of signals'' effect. Ref. \cite{CB2} took the absence of such distortion as the defining property of an exceptional theory and therefore asserted that Einstein-Gauss-Bonnet theory is not exceptional. However, this argument does not imply that Einstein-Gauss-Bonnet theory must suffer from shock formation since $\partial_\eta \beta$ never vanishes in this case. To demonstrate the possibility of shock formation, it is necessary to show that the theory has $N \ne 0$, as we will do below for Lovelock theories.
 
\section{Lovelock theories}

\subsection{Introduction}

In this section we will apply the general theory of the previous section to Lovelock theories. First we will demonstrate that the equation of motion of a Lovelock theory can be written in the form (\ref{evolution}). Then we will calculate the quantity $N$ defined by (\ref{Ndef}) and show that generically $N \ne 0$ so Lovelock theories are genuinely nonlinear, unlike GR.

\subsection{Lovelock theories are genuinely nonlinear}

We will consider Lovelock theories for which the coefficient of the Einstein term is non-vanishing. Normalizing this coefficient to $1$, the equation of motion (without matter) can be written\footnote{
Lower case Latin letters near the start of the alphabet ($a,b,c,\ldots $) are abstract indices, reserved for equations valid in any basis. Lower case Greek letters ($\alpha,\beta,\ldots, \mu,\nu,\ldots$) refer to a particular basis and take values $0,1,\ldots, (d-1)$. Lower case Latin letters from the middle of the alphabet ($i,j,k,\ldots$) take values $1, \ldots, (d-1)$.}
\be
\label{eqofmotiona}
0=E^a{}_b \equiv  \Lambda \delta^a_b+ G^a{}_b  + \sum_{p \ge 2} k_p \delta^{a c_1 \ldots c_{2p}}_{b d_1 \ldots d_{2p}} R_{c_1 c_2}{}^{d_1 d_2} \ldots R_{c_{2p-1} c_{2p}}{}^{d_{2p-1} d_{2p}}\;.
\ee
where $k_p$, $p\ge 2$ are the Lovelock coupling constants. The antisymmetry implies that only terms with $2p+1 \le d$ contribute. We can also write this as
\be\label{eqofmotionb}
 E^a{}_b \equiv \sum_{p \ge 0} k_p \delta^{a c_1 \ldots c_{2p}}_{b d_1 \ldots d_{2p}} R_{c_1 c_2}{}^{d_1 d_2} \ldots R_{c_{2p-1} c_{2p}}{}^{d_{2p-1} d_{2p}}\;,
\ee
where
\be
 k_0 = \Lambda \,,\qquad k_1 = -\frac14\,.
\ee
%\be
%\label{eqofmotion2}
% R_{ab} - \frac{2\Lambda}{d-2} g_{ab} + B_{ab} - \frac{1}{d-2} B^c{}_c g_{ab}  = 0\;.
%\ee
In a chart $x^\mu=(x^0,x^i)$ we have
\be
 R_{\mu\nu\rho\sigma} = \frac{1}{2} \left( g_{\mu\sigma,\nu\rho}  +g_{\nu\rho,\mu\sigma} -g_{\mu\rho,\nu\sigma} - g_{\nu\sigma,\mu\rho}  \right) + \ldots
\ee
The only Riemann components involving 2nd derivatives w.r.t. $x^0$ are
\be
\label{R0i0j}
 R_{0i0j} = -\frac{1}{2} \partial_0^2 g_{ij} + \ldots
\ee 
and components related to it by symmetry. Hence second $x^0$-derivatives of $g_{0\mu}$ do not appear in the equations of motion. We regard these metric components as non-dynamical. They have to be fixed by a gauge choice. For example if surfaces of constant $x^0$ are spacelike then $g_{0\mu}$ corresponds to the choice of lapse and shift. If a particular surface with constant $x^0$ is null then we could use Gaussian null coordinates, which also fix $g_{0\mu}$. 

We now define
 \be
 \label{A4def}
A^{\mu\nu\rho\sigma}=  -2 \sum_{p \ge 1} p k_p \delta^{0 \mu \rho i_3  \ldots i_{2p} | 0 \nu \sigma j_3 \ldots j_{2p}}  R_{i_3 i_4 j_3 j_4} \ldots R_{i_{2p-1} i_{2p} j_{2p-1} j_{2p}} \;,
\ee 
where $\delta^{\mu_1 \ldots \mu_n | \nu_1 \ldots \nu_n}$ is defined by raising the lower indices on $\delta^{\mu_1 \ldots \mu_n}_ { \rho_1 \ldots \rho_n}$. Using (\ref{R0i0j}), the terms involving 2nd derivatives w.r.t. $x^0$ in $E^{\mu \nu}$ can be written
\be
E^{\mu\nu} = A^{\mu\nu \rho\sigma} \partial_0^2 g_{\rho\sigma}+ \ldots = A^{\mu\nu i j} \partial_0^2 g_{ij}+\ldots
\ee
where the ellipses denotes terms independent of 2nd derivatives w.r.t. $x^0$. Note that 
\be
 A^{\mu\nu\rho\sigma} = \frac{\partial E^{\mu\nu}}{\partial(\partial_0^2 g_{\rho\sigma})} \equiv P(x,dx^0)^{\mu\nu\rho\sigma}\;,
\ee
where, for a 1-form $\xi_a$, $P(x,\xi)$ is the principal symbol of (\ref{eqofmotiona}). From this we deduce the basis-independent result\footnote{This differs from the principal symbol defined in \cite{us} because there the equation of motion was written in the ``trace-reversed'' form $R_{ab}  + \ldots=0$.}
\be\label{principalsymbol}
 P(x,\xi)^{abcd} = -2 \sum_{p \ge 1} p k_p \delta^{ac e f_3  \ldots f_{2p} | bde'  f_3' \ldots f_{2p}'}  \xi_e \xi_{e'} R_{f_3 f_4 f_3' f_4'} \ldots R_{f_{2p-1} f_{2p} f_{2p-1}' f_{2p}'} \;.
\ee 
Notice that this is symmetric on $ab$ and on $cd$, and
\be
 P(x,\xi)^{abcd} = P(x,\xi)^{cdab}
\ee
\be
\label{Pxi}
 P(x,\xi)^{abcd} \xi_a = P(x,\xi)^{abcd}\xi_c = 0\;.
\ee
Returning to our $(x^0,x^i)$ coordinate basis, this implies that
\be
 A^{\mu\nu\rho\sigma} = A^{\rho\sigma\mu\nu} \qquad A^{0\nu\rho\sigma} = A^{\mu\nu 0 \sigma} = 0\;.
\ee
The latter equality implies that $E^{0\mu}$ is independent of 2nd derivatives w.r.t. $x^0$ hence the equations $E^{0\mu}=0$ are constraints, just as in GR. 

The $ij$ components $E^{ij}=0$ are evolution equations. Note that $E^{ij}$ is {\it linear} in $\partial_0^2 g_{ij}$ and the coefficient of this term depends on second derivatives only of the form $\partial_i \partial_j g_{kl}$. Hence if we denote the dynamical fields $g_{ij}$ collectively as $g_I$ then the evolution equations take the form (\ref{evolution}) assumed in the previous section. The explicit $x^\mu$-dependence in (\ref{evolution}) arises through the dependence on the non-dynamical components $g_{0\mu}$. The terms involving 2nd $x^0$ derivatives in the evolution equations are $A^{ijkl} \partial_0^2 g_{kl}$. Hence in the notation of the previous section, $A_{IJ}$ corresponds to $A^{ijkl}$. 

Assume that the surface $x^0=0$ is characteristic, i.e., there exists non-zero (symmetric) $r_{ij}$ such that, at $x^0=0$,
\be
\label{Ardef}
 A^{ijkl} r_{kl} = 0\;.
\ee
From the symmetry of $A$ we have
\be
  r_{ij} A^{ijkl} = 0\;.
\ee
That is, the left and right eigenvectors of $A$ are the same ($l_I = r_I$ in the notation of the previous section). 

The symmetries of $A$ imply that (\ref{Ardef}) can be rewritten as
\be
 A^{\mu\nu\rho\sigma} r_{\rho\sigma} = 0\;.
\ee
Note that the components $r_{0\mu}$ do not contribute to this expression: they are pure gauge. More covariantly, a surface with normal $\xi_a$ is characteristic if, and only if, there exists symmetric $r_{ab}$ such that
\be
 P(x,\xi)^{abcd} r_{cd} = 0\;,
\ee
where $r_{cd}$ is not pure gauge, i.e., not of the form $\xi_{(c} X_{d)}$ for some $X_d$  \cite{us}. 

The coefficient of the nonlinear term in the transport equations for a discontinuity in second derivatives, or weak, high frequency waves, is $N$ defined by (\ref{Ndef}). Converting to the notation of this section gives
\be
N \equiv  \frac{\partial A^{ijkl}}{\partial (\partial_0 g_{mn})} r_{ij} r_{kl} r_{mn} \;.
 \ee
To calculate this we use\footnote{
If the characteristic hypersurface is non-null, as will be the case generically, then $\Gamma^0_{ij}$ is proportional to the extrinsic curvature of this surface.} 
\be
 \frac{\partial R_{ijkl}}{\partial (\partial_0 g_{mn})} r_{mn} = \Gamma^0_{i[k} r_{l]j} - \Gamma^0_{j[k} r_{l]i}\;.
\ee
We then find (see below Eq. (\ref{A4def}) for the definition of the Kronecker delta appearing here)
\be
\label{nonlinterm}
N
= -4 \sum_{p\ge 2} p(p-1) k_p \delta^{0ikmpr_5 \ldots r_{2p} | 0jlnp s_5 \ldots s_{2p}} \Gamma^0_{i j} r_{k l} r_{mn} r_{pq} R_{r_5 r_6 s_5 s_6} \ldots R_{r_{2p-1} r_{2p} s_{2p-1} s_{2p}}\;.
\ee
The sum starts at $p=2$ so (\ref{nonlinterm}) vanishes in GR, where $k_p=0$ for all $p \ge 2$. Hence GR is an exceptional theory. But if $k_p \ne 0$ for some $p \ge 2$ then the above expression does not vanish in general. This proves that Lovelock theories are genuinely nonlinear. 

\subsection{Shock formation in Lovelock theories}

We showed above that a Lovelock theory has an evolution equation of the form (\ref{evolution}) and so the derivation of the transport equations for a discontinuity in second derivatives of the metric, or weak high frequency gravitational waves, is a special case of the theory developed in sections \ref{discont} and \ref{waves}. 

First consider the case of a discontinuity in second derivatives of the metric, i.e., a discontinuity in curvature. This must propagate along a characteristic hypersurface. In coordinates $x^\mu=(x^0,x^i)$ adapted to this hypersurface we have
\be
 [\partial_0^2 g_{ij}] = \Pi(x) r_{ij}\;.
\ee
We should note that the components $[\partial_0^2 g_{0\mu}]$ are not gauge-invariant: they transform inhomogeneously under coordinate transformations which are $C^2$ but not $C^3$ at $x^0=0$ \cite{lich,anile}. However, the LHS above is gauge invariant and hence $\Pi$ is gauge invariant. 

We should briefly discuss the role of the constraint equations. Since these equations do not involve second derivatives w.r.t. $x^0$, they are continuous at $x^0=0$ and hence are satisfied automatically because the background solution (in the region $x^0<0$) satisfies them. So the constraints at $x^0=0$ do not impose any restrictions on $\Pi$. 

The behaviour of $\Pi$ along a bicharacteristic curve $x^i=x^i(s)$ within the surface $x^0=0$ is given by (\ref{Pisoln}). Shock formation corresponds to a vanishing denominator in this expression. As noted above, a shock will form if $\Pi(0)$ is large enough, with appropriate sign. We can estimate how large this must be by using dimensional analysis. Let's focus on the case of Einstein-Gauss-Bonnet theory, for which only the $p=2$ term is present in (\ref{nonlinterm}). 

Assume that components of the extrinsic curvature of the characteristic hypersurface are of order $R^{-1}$ for some length $R$ and adopt the convention that $r_{ij}$ is dimensionless. Then we have $N \sim k_2 /R$. Dimensional analysis suggests that $N(s) e^{-\Phi(s)}$ will not vary significantly for $0<s\ll R$ hence in this range the integral in (\ref{Pisoln}) is of order $k_2 s/R$. Hence the denominator in (\ref{Pisoln}) vanishes for $s \sim R/(k_2 \Pi(0))$. Self-consistency ($s \ll R$) requires $k_2 \Pi(0) \gg 1$. So a shock will form if the initial amplitude of the curvature discontinuity is large compared to the scale set by the Gauss-Bonnet coupling (and has appropriate sign). This is a sufficient condition for shock formation but not a necessary one. More generally, we could write the denominator of (\ref{Pisoln}) in the form $1+k_2 \Pi(0) f(s/R)$ and this will vanish at some positive value of $s$ when $k_2 \Pi(0)$ exceeds some critical value of order $1$. So a shock will form when the initial amplitude of the discontinuity is comparable to the scale set by the Gauss-Bonnet coupling (with appropriate sign). 

Now consider weak, high frequency, gravitational waves, as discussed in section \ref{waves}. Here we use the Ansatz 
\be
 g_{\mu\nu}(x,\eta) = \bar{g}_{\mu\nu}(x) + \omega^{-2} h_{\mu\nu}(x,\eta) + \omega^{-3} \kappa_{\mu\nu}(x,\eta) + \ldots
\ee
where $\eta = \omega x^0$ and $\bar{g}_{\mu\nu}$ is the ``background" solution. If we consider coordinate transformations of the form $x^\mu = \tilde{x}^\mu + \omega^{-3} \Psi^\mu(\tilde{x},\eta)$ then $h_{0\mu}$ transforms inhomogeneously but $h_{ij}$ is gauge invariant \cite{CBbook}. The analysis of section \ref{waves} shows that the surfaces of constant $x^0$ are characteristic hypersurfaces of the background spacetime and we can take
\be
 h_{ij}(x,\eta) = \Omega(x,\eta)  r_{ij}(x) 
\ee
for some function $\Omega(x,\eta)$, where $r_{ij}$ is defined using the principal symbol of the background solution. 

The analysis of section \ref{waves} involves only the evolution equations, not the constraints. In the Appendix, we show that the constraint equations do not impose any further restrictions on $\Omega$. 

The solution for $\Omega''$ is given in (\ref{Omegasol}). The same dimensional analysis argument that we used above shows that this blows up if $k_2 \Omega''(0,\alpha,\beta) \gg 1$ (with appropriate sign). Again, this is a sufficient condition for shock formation but not a necessary one. Generically we expect a shock to form when $k_2 \Omega''(0,\alpha,\beta)$ is of order 1 (with appropriate sign). Note that the high frequency waves make a contribution to the curvature of order $\Omega''$. Hence weak, high frequency waves, form a shock if the initial curvature is comparable compared to the scale set by the Gauss-Bonnet coupling (and has appropriate sign). 

These arguments indicate that shocks will form for a generic class of initial data with curvature comparable to the scale set by the Gauss-Bonnet coupling.  In some circumstances, shocks might form even when the curvature of the initial data is small compared to this scale. For example, in particular backgrounds, the integrals in (\ref{Pisoln}) or (\ref{Omegasol}) might grow monotonically with $s$. If this happens then a shock will form for arbitrarily small initial curvature. We will see an example of this when we discuss a plane wave spacetime below. 

Shock formation corresponds to divergent curvature (since $\Pi$ or $\Omega''$ diverges), and hence corresponds to the formation of a curvature singularity. This raises the question of whether this singularity is naked, or whether it is hidden inside a black hole. At the end of this paper we will argue that shocks are not always hidden inside black holes, i.e., weak cosmic censorship is violated in Lovelock theories without matter. Another question is whether the formation of a shock represents the end of time evolution, or whether one can develop a theory of the {\it evolution} of shocks, as is the case for a perfect fluid. This will also be discussed at the end of this paper.

Our discussion of shock formation presupposes that the initial value problem makes sense in Lovelock theories. But Ref. \cite{us} showed that Lovelock theories can fail to be hyperbolic when the curvature is comparable to the scale set by the coupling constants. This does not always happen, e.g. such theories are hyperbolic in any Ricci flat type N background, no matter how large the curvature may be \cite{us}. We need to check that the theory is hyperbolic for initial data that will form a shock. Consider the case of a  discontinuity propagating along a characteristic hypersurface $x^0=0$. On this hypersurface, the principal symbol is independent of the amplitude of the discontinuity (because it doesn't depend on $\partial_0^2 g_{\mu\nu}$). Hence the theory is hyperbolic on this hypersurface (assuming it is hyperbolic in the background spacetime). Therefore shock formation occurs in a region of spacetime in which the theory is hyperbolic so the initial value problem makes sense.

\subsection{Special cases with $N=0$}

Although $N$ is generically non-zero, it can still vanish under special circumstances. In this section we will show that $N$ vanishes for some of the simplest examples of characteristic surfaces in Lovelock theories. We will list four examples below, and then prove that $N$ vanishes for these cases. 

{\bf Example 1.} Flat spacetime. In this case, the principal symbol coincides with that of GR, so a hypersurface is characteristic if, and only if, it is null. It is not obvious that $N$ vanishes in this case because the $p=2$ term of (\ref{nonlinterm}) does not depend on the Riemann tensor. 

{\bf Example 2.} A Ricci flat spacetime with a Weyl tensor of type N in the classification of \cite{cmpp}. Any such spacetime is a solution of a Lovelock theory with $\Lambda=0$ \cite{CB2}. In such a spacetime, associated to the type N property is a preferred null vector field $\ell^a$. If this is hypersurface orthogonal then the null hypersurfaces orthogonal to it are characteristic \cite{us}. 

{\bf Example 3.} A spherically symmetric characteristic hypersurface in a static, spherically symmetric space-time. Such a surface must be null \cite{us}, i.e., ``gravity travels at the speed of light in the radial direction''. 

{\bf Example 4.} A Killing horizon. Ref. \cite{izumi} proved that a Killing horizon is a characteristic hypersurface in Einstein-Gauss-Bonnet theory. This result was generalized to an arbitrary Lovelock theory in Ref. \cite{us}. 

These examples have in common the feature that they all involve {\it null} characteristic surfaces. So first we'll discuss the case of a null characteristic surface $\Sigma$ in more detail. For such a surface we can introduce Gaussian null coordinates $(x^0,x^1,x^I)$ $(I=2,\ldots, d-1$) so that $\Sigma$ is given by $x^0=0$ and the metric in a neighbourhood of this surface is
\be
 ds^2 = 2 dx^0 dx^1  -{(x^0)}^2 F (d{x^1})^2 + 2x^0 h_I dx^1 dx^I + h_{IJ} dx^I dx^J\;.
\ee
where $F$, $h_I$ and $h_{IJ}$ depend on all the coordinates. 
This coordinate system is of the type discussed above; in these coordinates we have $x^i=(x^1,x^I)$. For most of the following, we will only need the metric evaluated at $x^0=0$:
\be
 ds^2|_{x^0=0} = 2 dx^0 dx^1 + h_{IJ} dx^I dx^J\;.
\ee
The condition for the surface to be characteristic is the existence of non-zero $r_{ij}$ satisfying (\ref{Ardef}). If we separate out the $p=1$ (GR) contribution from the $p \ge 2$ terms in this equation, and use the above metric, we obtain
\be
\label{nullchar}
 \delta_1^{(i} g^{j) k} r_{1k} - \frac{1}{2} \delta^i_1 \delta^j_1 g^{kl} r_{kl} - \frac{1}{2} g^{ij} r_{11} + B^{ijkl} r_{kl} = 0\;,
\ee
where
\be
B^{ijkl} = -2 \sum_{p \ge 2} p k_p \delta^{0 i k i_3  \ldots i_{2p} | 0 j l j_3 \ldots j_{2p}}  R_{i_3 i_4 j_3 j_4} \ldots R_{i_{2p-1} i_{2p} j_{2p-1} j_{2p}} \;.
\ee 
Let's now consider the examples discussed above. In flat spacetime we have $B^{ijkl}=0$. Solving the above equation then gives
\be
\label{transverse}
 r_{11} = r_{1I} = h^{IJ} r_{IJ} = 0\;.
\ee
In the case of a Ricci flat type N spacetime for which $\ell^a$ is hypersurface orthogonal, we choose our coordinates so that $\ell \propto dx^0$ which implies $\ell^\mu \propto \delta^\mu_1$. The only non-vanishing Riemann components are $R_{0J0K}$. But these do not contribute to $B^{ijkl}$. Hence $B^{ijkl}=0$ in this case so (\ref{transverse}) holds in this case too.  

For the example of a spherically symmetric hypersurface in a static, spherically symmetric spacetime, the coordinates $x^I$ parameterize a sphere $S^{d-2}$. In this case, the Riemann tensor components can all be written as functions of radius (of $S^{d-2}$) times products of $g_{AB}$ and $g_{IJ}$ where $A,B$ take values in $\{0,1\}$ \cite{us}. This implies that, at $x^0=0$, the only non-vanishing Riemann components are $R_{IJKL}$ and components $R_{\mu\nu\rho\sigma}$  with an equal number of $0$ and $1$ indices. Components with a (downstairs) $0$ index do not contribute to $B^{ijkl}$ so we have
\be
\label{specialB}
B^{ijkl} = -2 \sum_{p \ge 2} p k_p \delta^{0 i k I_3  \ldots I_{2p} | 0 j l J_3 \ldots J_{2p}}  R_{I_3 I_4 J_3 J_4} \ldots R_{I_{2p-1} I_{2p} J_{2p-1} J_{2p}} \;.
\ee 
Similarly, in the case of a Killing horizon, the fact that $\partial/\partial x^1$ is parallel to a null Killing vector field at $x^0=0$ implies  that $R_{1I1J} = R_{1IJK}=0$ at $x^0=0$ (see e.g. \cite{izumi,us}) so the only non-vanishing Riemann components of the form $R_{ijkl}$ are $R_{IJKL}$. Hence (\ref{specialB}) holds for this case too. So we will discuss our third and fourth examples together. At $x^0=0$ we have
\be
  \delta^{0 i k I_3  \ldots I_{2p} | 0 j l J_3 \ldots J_{2p}}= \delta^{0 i k I_3  \ldots I_{2p}}_{1 \rho\sigma J_3' \ldots J_{2p}'} g^{j\rho} g^{l \sigma} h^{J_3 J_3'} \ldots h^{J_{2p} J_{2p}'}\;,
 \ee
and for this to be non-zero we need either $i$ or $k$ to take the value $1$ (to balance the ``downstairs'' $1$) and either $\rho$ or $\sigma$ to take the value $0$ (to balance the ``upstairs'' $0$), which requires that either $j$ or $l$ takes the value $1$. So $B^{ijkl}$ is non-zero only if either $i$ or $k$ takes the value $1$ {\it and} either $j$ or $l$ takes the value $1$. Consider the $i=I$, $j=J$ components of (\ref{nullchar}). These give
\be
  \left( - \frac{1}{2} h^{IJ} + B^{IJ11} \right) r_{11} = 0\;,
\ee
and hence $r_{11}=0$. Now set $i=1$, $j=J$ in (\ref{nullchar}) to obtain
\be 
 \left( \frac{1}{2} h^{JK}  + 2 B^{1J1K} \right) r_{1K} = 0\;,
\ee
and hence $r_{1K}=0$. Finally set $i=1$, $j=1$ in (\ref{nullchar}) and use $r_{11}=0$ to obtain
\be
\label{Killinghor}
 \left( -\frac{1}{2} g^{KL} + B^{11KL} \right) r_{KL} = 0\;.
\ee
In the spherically symmetric case,  $B^{11KL} \propto h^{KL}$ so this equation implies $h^{KL} r_{KL} = 0$. Hence in this example, (\ref{nullchar}) is satisfied if, and only if, the conditions (\ref{transverse}) are satisfied. In the case of a Killing horizon, (\ref{nullchar}) is satisfied if, and only if, $r_{11} = r_{1I} =0$ and the condition (\ref{Killinghor}) is satisfied. 

In the first three examples, we have shown that the characteristic condition (\ref{Ardef}) reduces to the conditions (\ref{transverse}). These conditions are the same as the ``transverse'' condition for the polarization of a graviton in GR. They are $d$ conditions on the $d(d-1)/2$ components of $r_{ij}$. Similarly in the case of a Killing horizon we have $d$ independent conditions on $r_{ij}$. Hence in all four of our examples, there will be $d(d-1)/2-d=d(d-3)/2$ linearly independent solutions. This is the number of degrees of freedom of a graviton. Hence in all the examples, the null hypersurface is characteristic for {\it all} gravitational degrees of freedom. (For the case of a Killing horizon, this was proved in \cite{izumi,us}.) This is a consequence of the high degree of symmetry of these examples: generically, one expects only one solution for $r_{ij}$ for a given characteristic surface.\footnote{This expectation is confirmed by the results of Ref. \cite{us} for Ricci flat type N spacetimes, for the case of a generic characteristic hypersurface, i.e., one not orthogonal to $\ell_a$.}

To evaluate $N$ we first note that the only non-zero components of $\Gamma^0_{ij}$ on $\Sigma$ are
\be
 \Gamma^0_{IJ}|_{x^0=0} = -\frac{1}{2} \partial_1 h_{IJ}\;.
\ee
(The trace, and traceless part, of this are proportional to the expansion and shear of the generators of $\Sigma$.) Now $r_{11}=r_{1I}=0$ implies that the RHS of (\ref{nonlinterm}) reduces to
\be
 2 \sum_{p\ge 2} p(p-1) k_p \delta^{0I_1 \ldots I_4 i_5 \ldots i_{2p} | 0 J_1 \ldots J_4 j_5 \ldots j_{2p}} \partial_1 h_{I_1 J_1} r_{I_2 J_2} r_{I_3 J_3} r_{I_4 J_4} R_{i_5 i_6 j_5 j_6} \ldots R_{i_{2p-1} i_{2p} j_{2p-1} j_{2p}}\;.
\ee 
But at $x^0=0$,
\be
 \delta^{0I_1 \ldots I_4 i_5 \ldots i_{2p} | 0 J_1 \ldots J_4 j_5 \ldots j_{2p}} = \delta^{0I_1 \ldots I_4 i_5 \ldots i_{2p}}_{1 J'_1 \ldots J'_4 \rho_5 \ldots \rho_{2p}} h^{J_1 J_1'} \ldots h^{J_4 J_4'}  g^{j_5 \rho_5} \ldots g^{j_{2p} \rho_{2p}}\;,
\ee
and for the RHS of this to be non-zero we need one of the $i$ indices to be a $1$ and one of the $\rho$ indices to be a $0$.  Hence,  on the LHS, one of the $i$ indices must be a $1$ and one of the $j$ indices must be a $1$ (which is possible only for $p>2$ since otherwise there are no $i,j$ indices). But these indices are the ones contracted with the Riemann tensors. Therefore $N$ vanishes in flat spacetime. In our second example, the type N condition implies that any Riemann components with a downstairs $1$ index must vanish hence $N$ vanishes also in this case. In our third and fourth examples, any non-zero Riemann component with a downstairs $1$ index must also have a downstairs $0$ index. But none of the $i$ or $j$ indices can be zero because of the upstairs $0$s in the Kronecker deltas. Hence $N$ vanishes in these case too. 

We emphasize that the vanishing of $N$ in the above examples is atypical. It is a consequence of the special symmetries assumed in these examples. The most interesting generalization of these examples for which $N$ is non-zero would be to consider a {\it non-}spherically symmetric characteristic hypersurface in a static spherically symmetric spacetime. Ref. \cite{us} showed that a hypersurfaces is characteristic in such a background if, and only if, it is null w.r.t. one of three ``effective'' metrics, with the bicharacteristic curves corresponding to null geodesics of this effective metric. It would be interesting to pick a $(d-2)$-dimensional surface with axisymmetry, but not spherical symmetry, and determine the ``outgoing'' characteristic hypersurface emanating from it. It could be arranged that this hypersurface is free of caustics (unlike our plane wave example below with $N\neq0$). One could then study shock formation along such a hypersurface. 

We have ignored a technicality in the above discussion. In our derivation of the transport equations governing a curvature discontinuity or weak high frequency waves, we assumed that the eigenvector $r_{ij}$ satisfying (\ref{Ardef}) is non-degenerate, i.e., there exists a unique $r_{ij}$ on the characteristic surface in question. However, we have seen that this is not the case for the examples discussed above, for which there are $d(d-3)/2$ linearly independent  $r_{ij}$ obeying (\ref{Ardef}). The derivation of the transport equations can be generalized to allow for such degeneracy \cite{CB3,anile}. We will briefly describe the method here for the case of a curvature discontinuity. The treatment of weak high frequency waves is similar.

Following the notation of section \ref{discont}, denote by $l^{({\cal I})}_I$ and $r^{({\cal I})}_I$ a basis of solutions of (\ref{evolution2}), normalized in some convenient way. We then expand the discontinuity in terms of this basis as 
\be
[(g_J)_{00}] = \sum_{\cal I} \Pi_{\cal I} r^{({\cal I})}_J\;.
\ee
We then proceed as in section \ref{discont}, differentiating the equation of motion w.r.t. $x^0$, contracting with $l^{({\cal I})}_I$ and evaluating at $x^0=0$. This gives a system of ODEs for the quantities $\Pi_{\cal I}$. The nonlinear term in these ODEs is of the form $\sum_{\cal J,K} N_{\cal IJK} \Pi_{\cal J} \Pi_{\cal K}$ where $N_{\cal IJK}$ is obtained from (\ref{nonlinterm}) by the replacement of $r_{kl} r_{mn} r_{pq}$ by $r^{({\cal I})}_{kl} r^{({\cal J})}_{mn} r^{({\cal K})}_{pq}$. It is easy to see that the above argument generalizes immediately to this case: using $r^{\cal I}_{11} = r^{\cal I}_{1J}=0$, one finds that $N_{\cal IJK}=0$ for the examples discussed above so the transport equations are linear.

\subsection{Plane wave spacetime}

We saw above that $N$ vanishes in various simple situations. We emphasize that these examples are atypical and will now give an explicit example for which $N$ is non-zero. The spacetime is a homogeneous plane wave spacetime with metric
\be
 ds^2 = a_{IJ} x^I x^J du^2 + 2 du dv + \delta_{IJ} dx^I dx^J\;, \qquad a_{II} = 0\;.
\ee
This is a solution of any Lovelock theory with $\Lambda=0$ \cite{boulware}. It belongs to the class of Ricci flat type N spacetimes, whose characteristic hypersurfaces were determined in Ref.\cite{us}. Associated to the type N property is the null vector field $\ell^a = {\partial}/{\partial v}$ ($\ell_a = du$). 

\begin{figure}[htbp]
\centering
\includegraphics[width=8cm]{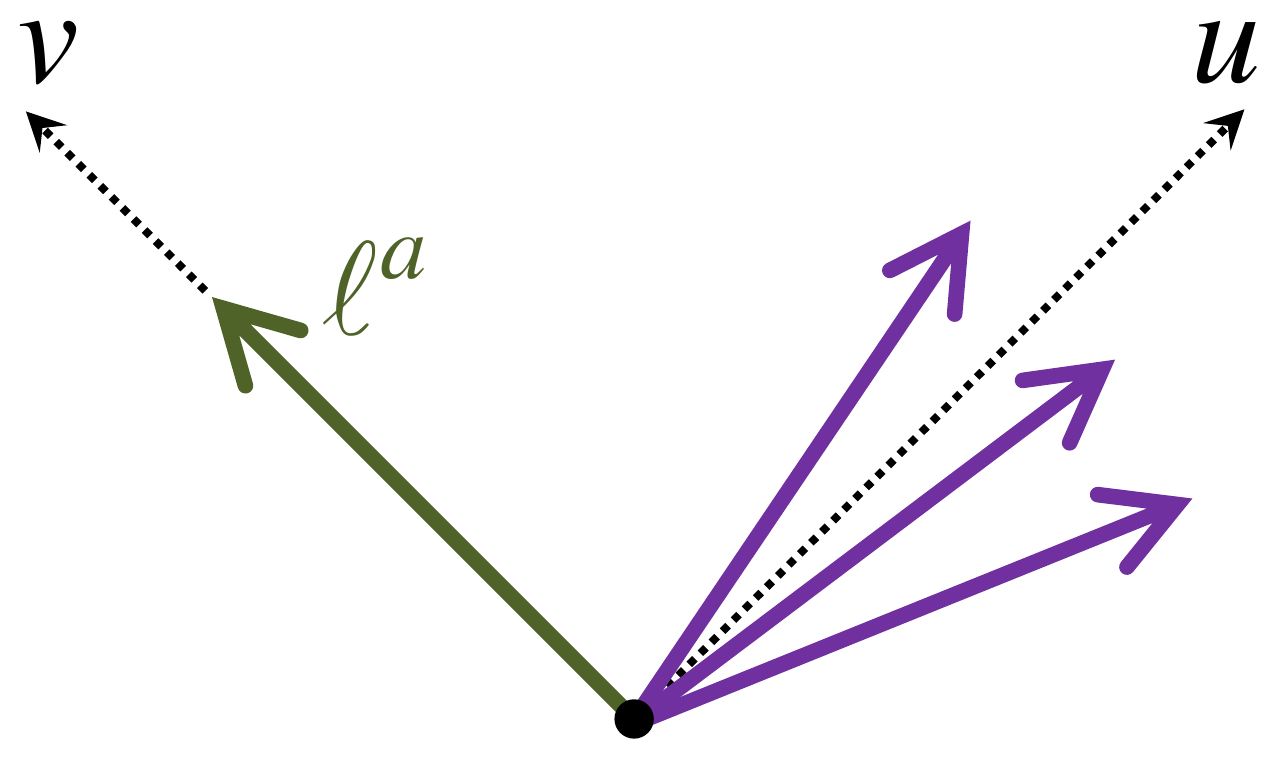}
\caption{Characteristic hypersurfaces in the plane wave spacetime. The coordinates $x^I$ are suppressed. The green and purple lines denote ``ingoing" and ``outgoing" characteristic hypersurfaces, respectively, emanating from the $(d-2)$-dimensional surface $u=v=0$.}
\label{Fig4}
\end{figure}

We will use results of Ref. \cite{us} to determine the characteristic hypersurfaces emanating from the surface $u=v=0$ in this spacetime. These fall into two classes (corresponding to the ``ingoing" and ``outgoing" families discussed in previous sections). One class consists of the null hypersurface $u=0$. Note that this has normal $\ell_a$ and hence is a special case of example 2 of the previous subsection. Hence it is characteristic for all gravitational degrees of freedom and has $N=0$. Therefore shocks do not form for disturbances propagating in the same direction as the plane wave itself. The second family is more complicated. Generically, there are $d(d-3)/2$ hypersurfaces in this family, one for each polarization of the graviton (see Fig.\ref{Fig4}). We will show that the ``outermost" one of these has $N \ne 0$, determine the transport equations for this hypersurface, and show that shocks form (before reaching a caustic) for {\it arbitrarily small} initial data. This effect can be attributed to focusing caused by the existence of a caustic on this hypersurface.

Ref. \cite{us} showed that for any Ricci flat type N spacetime, there exist $d(d-3)/2$ ``effective metrics" such that a hypersurface is characteristic if, and only if, it is null with respect to one of these metric.  Each effective metrics has the form
\be
 G_{ab} = g_{ab} - \omega l_a l_b\;,
\ee
where each $\omega$ depends on the spacetime curvature so $\omega$ is constant in the homogeneous plane wave spacetime. Hence we have
\bea
\label{effective}
 G_{\mu\nu} dx^\mu dx^\nu &=& (a_{IJ} x^I x^J - \omega ) du^2 + 2 du dv + \delta_{IJ} dx^I dx^J  \nonumber \\
 &=& a_{IJ} x^I x^J du^2 + 2 du dv' + \delta_{IJ} dx^I dx^J\;,
\eea
where
\be
 v' = v-\omega u /2\;.
\ee
This shows that, for this particular spacetime, each effective metric is isometric to the physical metric \cite{us}. The isometry depends on $\omega$ and is therefore different for each effective metric. 

To determine the characteristic hypersurfaces emanating from the surface $u=v=0$ we need to determine hypersurfaces which are null and orthogonal to this surface w.r.t. one of the effective metrics. This is equivalent to determining the null geodesics orthogonal to this surface w.r.t. each of the effective metrics. Since each of the effective metrics has the form (\ref{effective}), we just need to determine the null geodesics of this metric that are orthogonal to the surface $u=v'=0$. 

Let $\lambda$ be an affine parameter along such a geodesic with $\lambda=0$ on this surface. Orthogonality implies that the tangent vector at $\lambda=0$ must be a linear combination of $\partial/\partial u$ and $\partial/\partial v'=\partial/\partial v$. There are two possibilities (up to scaling): $\partial/\partial u - (1/2) a_{IJ} x^I x^J \partial/\partial v'$ and $\partial/\partial v'=\partial/\partial v$. The latter corresponds to the trivial case ($u=0$ hypersurface) discussed above so let us focus on the former.

The geodesic equation for $G_{ab}$ gives (dot denotes differentiation by the affine parameter $\lambda$)
\be
 \dot{u} = P \qquad \Rightarrow \qquad u(\lambda) = P \lambda\;,
\ee
for some constant $P$. $P$ must be non-zero because we know the tangent vector at $\lambda=0$ has a non-vanishing $u$-component. We normalize the affine parameter so that $P=1$ so
\be
 u(\lambda) = \lambda\;.
\ee
The geodesic equation for $x^I$ gives
\be
 \ddot{x}^I -  a_{IJ} x^J = 0\;.
\ee
Without loss of generality, we can assume that $a_{IJ}$ is diagonal with components $a_I$ and henceforth we will not use the summation convention for indices $I,J,\ldots$. Using the form of the tangent vector at $\lambda=0$, the solution of this equation is
\be
\label{geod}
 x^I(\lambda) = \eta^I \cosh (\sqrt{a_I}  \lambda)\;,
\ee
where $\eta^I =x^I(0)$. This solution is valid for $a_I \le 0$ as well as $a_I>0$. The condition that the geodesic is null w.r.t. $G_{ab}$ gives an equation for $\dot{v}'$ which can be integrated to give
\be
 v'(\lambda) = -\frac{1}{4} \sum_I \sqrt{a_I} \eta^{I2} \sinh( 2 \sqrt{a_I}  \lambda) \;.
\ee
In terms of the original coordinates we have
\be
u=\lambda \qquad x^I = \eta^I \cosh (\sqrt{a_I} \lambda) \qquad v = \frac{\omega}{2} \lambda -  \frac{1}{4} \sum_I \sqrt{a_I} \eta^{I2} \sinh( 2 \sqrt{a_I} \lambda) \;.
\ee
This defines our characteristic hypersurface with parameters $(\lambda,\eta^I)$. The curves of constant $\eta^I$ are the bicharacteristic curves within this hypersurface. These curves are generically non-null w.r.t. the physical metric. 

Since $\sum_I a_{I} = 0$, at least one of the $a_{I}$ must be negative. Assume $a_{1}<0$ and consider bicharacteristic curves with $\eta^2=\eta^3= \cdots 0$. All such curves intersect when $\lambda \sqrt{a_I} = \pm i\pi/2$, i.e., $\lambda = \pm \pi/(2\sqrt{-a_1})$. Hence our characteristic surface contains caustics.

Let us switch to a coordinate system adapted to our characteristic hypersurface.  Let
\be
\eta^I=\frac{x^I}{\cosh(\sqrt{a_I} u)}\;,\qquad x^0=v-\frac{\omega}{2} u+ \frac{1}{4} \sum_I \sqrt{a_I} \eta^{I2} \sinh( 2 \sqrt{a_I} u)\;,
\ee
so that $x^0=0$ is our characteristic surface.  Then the physical metric becomes\footnote{
A coordinate change $x^0=w-\omega u/2$ gives the ``Rosen form" of the plane wave metric.}
\be
ds^2=2dx^0 du+\omega du^2+\sum_I \cosh^2(\sqrt{a_I}u)(d\eta^I)^2\;.
\ee
Note that these coordinates break down at a caustic. We will denote the coordinates $(u,\eta^I)$ collectively by $x^i$, $i=1,\ldots, d-1$, as in previous sections. Note that the characteristic surface is spacelike if $\omega>0$ and timelike if $\omega<0$. 

The allowed values of the constant $\omega$ are determined by imposing the condition that the above surface is characteristic. To do this, we must find a symmetric tensor $r_{\mu\nu}$ that is in the kernel of the principal symbol $P(x,dx^0)$ given in \eqref{principalsymbol}, which is not pure gauge (i.e. $r_{ij} \ne 0$).  This was done in Ref. \cite{us} for a general Ricci flat type N spacetime.\footnote{Note that even though the principal symbol in \cite{us} uses the trace-reduced equations of motion, and so is different than \eqref{principalsymbol}, the right eigenvectors $r_{\mu\nu}$ are unchanged.} We summarize the results here. 

The allowed values of $\omega$ are of two types. They depend only on the Gauss-Bonnet coupling $k_2$ and not on $k_p$ for $p>2$. The first type is given for $I \ne J$ by
\be
\omega = \omega_{\{I,J\}} \equiv 32 k_2 (a_I+a_J)
\ee 
and the associated right eigenvector $r_{\mu\nu}$ is given by
\be
  {r_{\{I,J\}}}^\mu{}_\nu= \left\{
     \begin{array}{ll}
       1 & : \{\mu,\nu\}=\{I,J\}\\
       0 & : \mathrm{otherwise}
     \end{array}
   \right.\;,\qquad (I\neq J)\;.
\ee
This gives $(1/2)(d-2)(d-3)$ values of $\omega$. The remaining $d-3$ values are obtained by solving an eigenvalue problem for a $(d-3) \times (d-3)$ symmetric matrix. For $d=5$ we can give the result explicitly. The allowed values of $\omega$ in this case are 
\be
 \omega = \omega_\pm \equiv \mp32k_2\nu\;,
\ee
where
\be
\nu=\sqrt{\frac{2}{3}(a_1^2+a_2^2+a_3^2)}\;.
\ee
The associated $r_{\mu\nu}$ are given by
\be
r_\pm{}^\mu{}_\nu=\mathrm{diag}\left[\mp\frac{1}{\nu},\mp\frac{1}{\nu},
\frac{3(2a_1\mp\nu)}{2(a_1-a_2)(a_1-a_3)},
\frac{3(2a_2\mp\nu)}{2(a_2-a_1)(a_2-a_3)},
\frac{3(2a_3\mp\nu)}{2(a_3-a_1)(a_3-a_2)}
\right]\;.
\ee
Note that the components $r_{0\mu}$ are pure gauge. We have made a particular choice of gauge and normalization in the above expressions for $r_{\mu\nu}$. 

Ref. \cite{us} showed that, at any point of a Ricci flat type N spacetime, the light-cones of the effective metrics form a nested set. Causality is determined by the effective metric with the outermost light cone. This corresponds to the effective metric with the most positive value of $\omega$, i.e., the ``most spacelike" characteristic surface. For the above spacetime with $d=5$ this corresponds to $\omega=\omega_-$. Hence if we are interested in a discontinuity ``invading" a background plane wave spacetime, then $\omega=\omega_-$ will correspond to the outermost characteristic surface, which separates the background spacetime from the spacetime on the other side of the discontinuity (see section \ref{discont}). 

Now we can determine $N$, the nonlinear term in the transport equations. Henceforth we will assume $d=5$ so that we can use the above expressions. For $d=5$, only the $p=2$ (Gauss-Bonnet) term is present in (\ref{nonlinterm}). This gives
\be
N
= -8  k_2 \delta^{0ikmp | 0jlnp } \Gamma^0_{i j} r_{k l} r_{mn} r_{pq}\;.
\ee
The non-zero components of $\Gamma^0_{ij}$ are 
\be
\Gamma^0_{II}=-\frac{1}{2}\sqrt{a_I}\sinh(2\sqrt{a_I}u)\;.
\ee
We can now evaluate $N$ for the characteristic hypersurfaces corresponding to different values of $\omega$. First, for $\omega=\omega_{\{I,J\}}$ we find that setting $r=r_{\{I,J\}}$ gives $N=0$ since $r_{\{I,J\}}$ does not have enough nonzero components. So there is no shock formation along these hypersurfaces. Now consider $\omega=\omega_\pm$. Setting $r=r_\pm$ we find the non-zero result
\begin{align}\label{planewaveN}
 N&=\mp\frac{108 k_2\omega_\pm}{\nu} \Big[\frac{(2a_2\mp\nu)(2a_3\mp\nu)t_1}{(a_1-a_2)(a_1-a_3)(a_2-a_3)^2}\nonumber\\&\qquad+\frac{(2a_1\mp\nu)(2a_3\mp\nu)t_2}{(a_2-a_1)(a_2-a_3)(a_1-a_3)^2}+\frac{(2a_1\mp\nu)(2a_2\mp\nu)t_3}{(a_3-a_1)(a_3-a_2)(a_1-a_2)^2}\Big]\;,
\end{align}
where
\be \label{tdef}
t_I=\sqrt{a_I}\tanh(\sqrt{a_I}u)\;.
\ee
Note that $N$ diverges at a caustic.

To determine the full transport equations we could use the general results given in sections \ref{discont} and \ref{waves}. But the linear term is quite complicated so it is easier to rederive the equations using computer algebra as follows. First consider the case of a curvature discontinuity propagating along the characteristic hypersurface $x^0=0$. For $x^0<0$ our physical metric is the plane wave metric give above. But for $x^0>0$ it is different, with a discontinuity in second derivatives at $x^0=0$:
\be
 [\partial_0^2 g_{\mu\nu}] = \Pi r_{\mu\nu}\;.
\ee
The general theory presented in section \ref{discont} shows that we can derive an equation which depends only on this discontinuity and the background solution. Hence we can derive this equation by writing down an Ansatz for the metric in $x^0>0$ which has the correct discontinuity:
\be
g_{\mu\nu}=\bar g_{\mu\nu}+\frac{1}{2}(x^0)^2\Pi(u,\eta^I)r_{\mu\nu} \qquad x^0>0\;,
\ee
where $\bar{g}_{\mu\nu}$ is the background solution and we take $r=r_\pm$. The above metric will not solve the equations of motion in $x^0>0$ but it will give the correct evolution equation for the discontinuity at $x^0=0$. We now follow, using computer algebra, the steps of section \ref{discont} to obtain the evolution equation for $\Pi$. The result is an equation of the form (\ref{mastereq1}) with $N$ given above, 
\be\label{planewaveKi}
K^i= K \delta^{i}_u\;,
\ee
where $K$ is a constant given by
\be\label{planewaveK}
K=-\frac{9}{4}\left(\frac{(2a_1\mp\nu)^2}{(a_1-a_2)^2(a_1-a_3)^2}+\frac{(2a_2\mp\nu)^2}{(a_2-a_1)^2(a_2-a_3)^2}+\frac{(2a_3\mp\nu)^2}{(a_1-a_3)^2(a_2-a_3)^2}\right)\;,
\ee
and
\be
M= \frac{K}{2}(t_1+t_2+t_3)\;,
\ee
where $t_I$ is given in \eqref{tdef}. Note that $K^i \propto \delta^i_u$ implies that the integral curves of $K^i$ are the bicharacteristic curves, as expected. In terms of the parameter $s$ along these curves defined by (\ref{sdef}) we have
\be
u=Ks\;.
\ee

A similar procedure can be used to obtain the transport equation for weak, high frequency, gravitational waves. In this case we put
\be
g_{\mu\nu}=\bar g_{\mu\nu}+\omega^{-2}\Omega(x^0,u,\eta^I,\eta)r_{\mu\nu}
\ee
into the equations of motion, where $\eta=\omega x^0$, and evaluate the equations of motion to first order in $\omega^{-1}$ and contract with $r_{\mu\nu}$.  The result is a transport equation of the form \eqref{OmegaPDE} with $K^i$ and $N$ as given above and
\be
\tilde{M}=M\;.
\ee

We can now discuss shock formation in the plane wave background. Let us focus on the case of a discontinuity in curvature, for which the solution along a bicharacteristic curve is given in (\ref{Pisoln}). We find that
\be
 e^{-\Phi(s)} = \Pi_{I=1}^3 \left( \cosh ( \sqrt{a_I} K s)\right)^{-1/2}\;.
\ee
For definiteness, let's focus on the case for which $a_1<0$ and $a_2,a_3>0$. Then we have a caustic at $u=- \pi/(2\sqrt{-a_1})$ and hence at $s = \pi/(2|K|\sqrt{-a_1}) \equiv s_*$ (note that $K<0$). As we approach $s_*$ we have
\be
 N e^{-\Phi} \propto (s_*-s)^{-3/2}
\ee
and hence the integral in (\ref{Pisoln}) diverges as $(s_*-s)^{-1/2}$ as $s \rightarrow s_*-$. As long as $\Pi(0)$ has the right sign, the denominator in (\ref{Pisoln}) will vanish at $s=s_0$ for some $0<s_0<s_*$. This implies that $\Pi(s)$ diverges as $(s_0-s)^{-1}$ as $s \rightarrow s_0-$, corresponding to shock formation. Note that this occurs for {\it arbitrarily small} $\Pi(0)$. Similar results hold for weak, high frequency waves. 

It is easy to understand why shock formation occurs for small initial data here. The presence of the caustic focuses the initial discontinuity. This can be seen by considering the analogous problem in GR e.g. by setting $k_2=0$ above. The solution (\ref{Pisoln}) reduces to $\Pi(s) = \Pi(0) e^{-\Phi(s)}$, which diverges as $(s_*-s)^{-1/2}$. So even in GR one has a divergence, at the caustic, caused by focusing.\footnote{
This is sometimes called a linear shock because it occurs even for linear equations \cite{anile}.}
 In Einstein-Gauss-Bonnet theory, this focusing causes the amplitude of the discontinuity to grow as the caustic is approached. We have argued above that, generically, a shock must form once the amplitude is large enough (and the sign is right) and indeed a shock does form {\it before} reaching the caustic. 

\section{Discussion}
 
 \subsection{Shock formation from smooth initial data}
 
We have shown that Lovelock theories are genuinely nonlinear and hence suffer from divergences analogous to the formation of shocks in a perfect fluid. We have shown this by considering (i) solutions with a curvature discontinuity and (ii) weak high frequency gravitational waves. In both cases, a divergence occurs in finite time whenever the initial amplitude of the disturbance is large enough (on a scale set by the Lovelock coupling constants) with appropriate sign. 

Based on what is known for other genuinely nonlinear theories, it seems likely that shocks will also form in exact solutions arising from a large class of {\it smooth} initial data. It would be interesting to find explicit examples of this. As discussed in the Introduction, many genuinely nonlinear theories admit plane wave solutions with arbitrarily small initial amplitude that blow up in finite time \cite{john}. Of course, Lovelock theories also admit plane wave solutions: any Ricci flat pp-wave solution of GR is also a solution of any Lovelock theory with $\Lambda=0$ \cite{boulware}. But such solutions do not blow up. This appears closely related to our result that weak high frequency gravitational waves never form shocks in a flat background spacetime, even though they can form shocks in a generic background spacetime. Hence presumably the behaviour of pp-wave solutions is not typical of the behaviour of more general solutions. 

\subsection{Weak cosmic censorship}

If we start from geodesically complete, asymptotically flat, initial data that forms a shock then is this singularity naked, or does it occur in the interior of a black hole? We have argued that shocks will form for {\it outgoing} disturbances if these are strong enough. This does not seem related to the usual mechanisms for black hole formation, namely gravitational collapse or focusing of ingoing gravitational waves. Furthermore, if the amplitude of the initial disturbance is decreased, then the ``time'' it takes for the shock to form increases. Hence the wavefront at the time of shock formation is likely to be larger for a weak initial disturbance than for a strong one. If the shock is to be hidden inside a black hole, this implies that the black hole would have to be larger for a weak initial disturbance than for a strong one, which seems unlikely because the weaker disturbance would have smaller energy. This suggests that shocks are not always hidden inside black holes.

We have not been very careful with our use of the term ``black hole" in the above paragraph. This term is ambiguous in Lovelock theories because the causal structure is not determined by the light cone. A better way of posing the question is to ask whether any signal can be sent from the shock to future null infinity, i.e., whether there exists a bicharacteristic curve extending from the shock to future null infinity.\footnote{An even better formulation is to ask whether the ``maximal development" of such initial data is (generically) an asymptotically flat spacetime with a complete future null infinity.}

In the case of a curvature discontinuity ``invading" an asymptotically flat background spacetime, it is clear that a signal can be sent from the shock to future null infinity. To see this, note that characteristic hypersurfaces of the background spacetime approach null hypersurfaces near infinity (because the Lovelock terms are negligible when the curvature is small). Hence such hypersurfaces intersect future null infinity. Given an initial $(d-2)$-dimensional surface $S$ of spherical topology, pick an ``outermost" outgoing characteristic hypersurface $\Sigma$ emanating from $S$. Assume no caustic forms on $\Sigma$. Then we can arrange a shock to form on $\Sigma$ by taking the initial amplitude of the discontinuity to be large enough. This shock is ``visible" to future null infinity because $\Sigma$ extends to future null infinity. This suggests that the same will happen for a shock that forms from smooth initial data.

In summary, it seems likely that shock formation implies that weak cosmic censorship is violated in Lovelock theories without matter. This discussion assumes that one cannot evolve the solution further once a shock forms. However, if it is possible to develop a theory of the {\it evolution} of shocks (see below) then this would enlarge the class of admissible spacetimes to allow for dynamical shocks, and shock formation might be consistent with a version of weak cosmic censorship in this enlarged class of spacetimes. 

\subsection{Nonlinear stability of Minkowski spacetime}

Genuinely nonlinear theories can form shocks if an initial disturbance is large enough. What about ``small" initial data, i.e., initial data close to some trivial solution? In some cases, this can also lead to shock formation. For example, consider a (compressible) perfect fluid in 3+1 dimensions. In this case, it has been proved that small initial data of compact support can form shocks \cite{sideris,christodouloufluids}. 

Small initial data in a Lovelock theory with $\Lambda=0$ corresponds to almost flat initial data. The formation of a shock starting from such initial data would correspond to a nonlinear instability of Minkowski spacetime.  We will argue that Minkowski spacetime is {\it stable} in Lovelock theories, essentially because such theories are higher dimensional. 

In harmonic coordinates, the equation of motion of a Lovelock theory takes the form
\be
 \Box h_{\mu\nu} = {\cal F}_{\mu\nu}(h,\partial h,\partial^2 h)\;,
\ee 
where $h_{\mu\nu} = g_{\mu\nu} -\eta_{\mu\nu}$, $\Box$ is the Minkowski spacetime wave operator and the RHS is of quadratic order. Compare this with a nonlinear scalar wave equation in Minkowski spacetime
\be
 \Box \phi = {\cal F}(\phi,\partial \phi,\partial^2 \phi)\;,
\ee
where ${\cal F}$ is of quadratic order. In this case, for $d>4$, it is known that the trivial solution $\phi=0$ is stable: any solution arising from small amplitude, compactly supported, initial data will decay \cite{hormander}.\footnote{
For $d=5$ this requires the extra condition that ${\cal F}(\phi,0,0) = {\cal O}(\phi^3)$, which would be satisfied in the analogy with a Lovelock theory, for which ${\cal F}_{\mu\nu}(h,0,0)=0$.} For $d=4$, the problem is much harder because the slower decay of solutions of the linearized equation of motion make it harder to control the nonlinear terms. For $d>4$, solutions of the linearized equation decay faster because there are more dimensions for a disturbance to spread into.

This analogy suggests that Minkowski spacetime is stable in Lovelock theories, essentially because the higher-dimensional nature of such theories guarantees that solutions of the linearized equation decay sufficiently rapidly that nonlinear effects do not become important. This is the same reason why proving stability of Minkowski spacetime in GR is expected to be much easier in higher dimensions than the four-dimensional case. In the $d=4$ case the proof is highly non-trivial \cite{minkstable}.

The above discussion assumed asymptotically flat boundary conditions. But one could also consider Kaluza-Klein boundary conditions, with $d-4$ compact dimensions. In this case, it seems plausible that Lovelock theories would behave analogously to a perfect fluid in $3+1$ dimensions, with blow-up for small initial data, i.e., flat spacetime would be unstable with these boundary conditions.

\subsection{Evolution of shocks}

We have used the word ``shock" in this paper because the mechanism behind singularity formation appears to be the same as for a compressible perfect fluid. In fluid mechanics, the formation of a shock does not represent the end of time evolution: there is a theory governing the evolution of shocks. This theory is based on the notion of {\it weak} solutions to the equations of motion. Once a shock forms, one continues the solution by allowing the fluid variables to be discontinuous across a hypersurface (the shock). For a perfect fluid, conservation of energy-momentum and particle number leads to a set of junction conditions (the Rankine-Hugoniot conditions) which connect the solutions on the two sides of the shock. The shock propagates along a {\it non}-characteristic hypersurface. This hypersurface travels faster than sound w.r.t. the fluid outside the shock, and slower than sound w.r.t. the fluid inside.

Could one do something similar for Lovelock theories? The analogous procedure appears to be to consider a hypersurface across which the {\it first} derivative of the metric (extrinsic curvature) is discontinuous. A natural notion of weak solution is to demand that the fields extremize the action even in the presence of the discontinuity.\footnote{
If one does this in GR then one finds that the discontinuity must propagate along a characteristic (i.e. null) hypersurface \cite{taub}, which does not correspond to a shock but simply to propagation of a feature already present in the initial data.}  This is the same way that the junction conditions for a domain wall are derived so the result is the same as these junction conditions, but with no matter source term present. In adapted coordinates $(x^0,x^i)$ so that the hypersurface is at $x^0=0$, the junction condition specifies the discontinuity in the the canonical momentum $\pi^{ij}$ conjugate to the metric components $g_{ij}$ \cite{barrabes}. Hence the junction condition with no matter source is that $\pi^{ij}$ should be continuous at $x^0=0$. 

In a Lovelock theory, $\pi^{ij}$ is a non-linear polynomial in the extrinsic curvature of the surface \cite{teitelboim}. Hence, unlike in GR, it is possible for $\pi^{ij}$ to be continuous even if the extrinsic curvature is not. This suggests that it might be possible to define a shock in a Lovelock theory as a hypersurface $\Sigma$ across which the extrinsic curvature is discontinuous but $\pi^{ij}$ must be continuous.\footnote{Hypersurfaces satisfying such junction conditions have been discussed previously, with different motivation (see e.g. \cite{Garraffo:2007fi}).} In analogy with a perfect fluid, it might be necessary to demand that this surface travel ``faster than gravity" w.r.t. the spacetime on one side of the shock, and ``slower than gravity" w.r.t. the spacetime on the other side. More precisely, consider an outgoing shock in an asymptotically flat spacetime. The shock front should ``catch up with" outgoing characteristic hypersurfaces outside the shock. Inside the shock, the outgoing characteristic hypersurfaces should catch up with the shock.  

\begin{figure}[htbp]
\centering
\includegraphics[width=12cm]{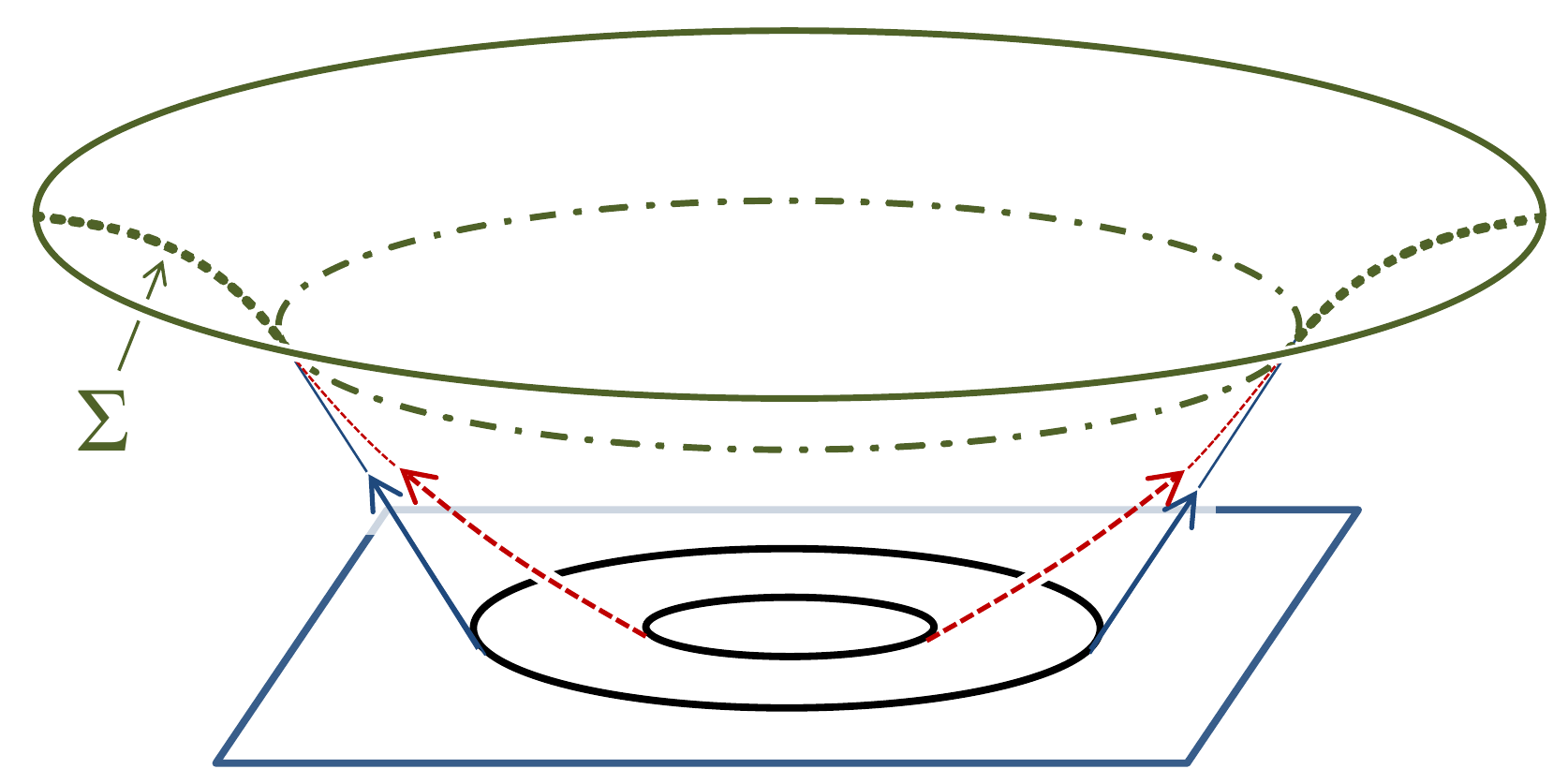}
\caption{Shock evolution. Starting from smooth initial data, as in Fig. \ref{Fig2}, a shock forms on the dot-dashed line. The solution is extended by allowing first derivatives of the metric to be discontinuous across a non-characteristic hypersurface (in green) satisfying the junction condition that $\pi^{ij}$ should be continuous.}
\label{Fig5}
\end{figure}

Shock formation and evolution in Lovelock theories might be treated following the discussion for a perfect fluid in Ref. \cite{christodouloufluids}. Consider smooth initial data which leads to a solution that blows up on a $(d-2)$-dimensional surface. Now try to extend the solution further by allowing the first derivative of the metric to be discontinuous across a hypersurface $\Sigma$ emanating from this $(d-2)$-dimensional surface, demanding continuity of $\pi^{ij}$ across $\Sigma$. See Fig. \ref{Fig5}. It would be interesting to see whether this can be done.  

\subsection*{Acknowledgments}

We are grateful to M. Dafermos, J. Keir and J. Luk for useful discussions. This work was supported by the European Research Council grant no. ERC-2011-StG 279363-HiDGR. N.T.\ was supported in part by the World Premier International Research Center Initiative (WPI Initiative), MEXT, Japan, and by JSPS Grant-in-Aid for Scientific Research 25$\cdot$755.

\appendix
\section{Constraint equations for gravitational waves}
Lovelock theories have additional structure to that of the theories described in section 2, namely gauge freedom and constraints. Here, we will show that these do not affect the transport equations.  The starting point is an expansion
\be
 g_{\mu\nu}(x) = \bar{g}_{\mu\nu}(x) + {\omega}^{-2} h_{\mu\nu}(x,\eta) + \omega^{-3} \kappa_{\mu\nu}(x,\eta) + \ldots\;,
\ee 
%with $\eta = \omega x^0$. In the notation of section \ref{waves}, $g_I$ corresponds to $g_{ij}$. We can consider coordinate transformations of the form $x^{\mu} \rightarrow x^\mu + \omega^{-3} \psi^\mu(x,\eta)$. The result is \cite{CBbook}
%\be
 %h_{\mu\nu} \rightarrow h_{\mu\nu} +2  \bar{g}_{\rho(\mu } \psi'^\rho \delta^0_{\nu)} 
%\ee
%and hence the components $h_{0\mu}$ are gauge-dependent but the components $h_{ij}$ are gauge-invariant. These are the components corresponding to $h_I$ in section \ref{discont}. So the transport equation derived there involves only gauge-invariant quantities.
%The constraints in this case can be understood as for weak high frequency waves in GR \cite{CBbook}. The constraint equations are non-trivial at ${\cal O}(\omega^{-1})$. However, these equations involve the gauge-dependent components $h_{0\mu}$ {\bf check detailsÉ}
where $\eta=\omega \phi(x)$, and zero set of $\phi$ is the surface of constant phase for the waves. (Previously we choose coordinates so that $\phi=x^0$, we will not do that here.) Working in a coordinate basis, the Riemann tensor to $O(\omega^{-1})$ is given by \cite{CBbook}
\be
R_{\mu\nu}{}^{\rho\sigma}=\bar R_{\mu\nu}{}^{\rho\sigma}+S_{\mu\nu}{}^{\rho\sigma}+\omega^{-1}\, T_{\mu\nu}{}^{\rho\sigma}+O(\omega^{-2})\;,
\ee
where $n_\mu=\partial_\mu\phi$, $\bar R_{\mu\nu}{}^{\rho\sigma}$ is the Riemann tensor for the background $\bar g$, and
\begin{align}
S_{\mu\nu}{}^{\rho\sigma}&=2n_{[\mu}h''_{\nu]}{}^{[\rho}n^{\sigma]}\\
T_{\mu\nu}{}^{\rho\sigma}&=2(\bar\nabla^{[\rho}h'^{\sigma]}{}_{[\mu}n_{\nu]}+\bar\nabla_{[\mu}h'_{\nu]}{}^{[\rho}n^{\sigma]}-h'_{[\mu}{}^{[\rho}\bar\nabla_{\nu]}n^{\sigma]}+n_{[\mu}\kappa''_{\nu]}{}^{[\rho}n^{\sigma]})\;.
\end{align}
Indices are raised and lowered by the background metric $\bar g$.  We then expand the equations of motion in inverse powers or $\omega$.  At lowest order, we find
\be
E^\mu{}_\nu=\bar E^\mu{}_\nu-2\sum_{p \ge 0} p k_p \delta^{\mu \sigma_1 \ldots \sigma_{2p}}_{\nu \rho_1 \ldots \rho_{2p}}n_{\sigma_1}n^{\rho_1}h''_{\sigma_2}{}^{\rho_2}\bar R_{\sigma_3 \sigma_4}{}^{\rho_3 \rho_4} \ldots \bar R_{\sigma_{2p-1} \sigma_{2p}}{}^{\rho_{2p-1} \rho_{2p}}+O(\omega^{-1})\;.
\ee
Following the discussion in section \ref{waves}, if $\bar g$ is a background solution, then $\bar E^\mu{}_\nu=0$, and $h$ can be integrated to give something of the form $h_{\mu\nu}(x,\eta)=\Omega(x,\eta) r_{\mu\nu}(x)$, where $r_{\mu\nu}$ is in the kernel of the principal symbol \eqref{principalsymbol}.

The transport equations are given by the next order $O(\omega^{-1}$) in the expansion.  We will now show that the constraints equations (given by $n_\mu E^\mu{}_\nu=0$) are automatically satisfied at this order.  The contraction with $n$ ensures that any terms involving $S_{\mu\nu}{}^{\rho\sigma}$ would vanish by antisymmetry.  Therefore, we are left with terms involving just $T_{\mu\nu}{}^{\rho\sigma}$ and $\bar R_{\mu\nu}{}^{\rho\sigma}$.  Further terms in $T_{\mu\nu}{}^{\rho\sigma}$ drop out due to antisymmetry and we are left with
\begin{align}
n_\mu E^\mu{}_\nu&=\sum_{p \ge 0} p k_p \delta^{\mu \sigma_1 \ldots \sigma_{2p}}_{\nu \rho_1 \ldots \rho_{2p}}(\bar\nabla_{\sigma_1}h'_{\sigma_2}{}^{\rho_1}n^{\rho_2}-h'_{\sigma_1}{}^{\rho_1}\bar\nabla_{\sigma_2}n^{\rho_2})n_\mu\bar R_{\sigma_3 \sigma_4}{}^{\rho_3 \rho_4} \ldots \bar R_{\sigma_{2p-1} \sigma_{2p}}{}^{\rho_{2p-1} \rho_{2p}}+O(\omega^{-2})\nonumber\\
&=\sum_{p \ge 0} p k_p \delta^{\mu \sigma_1 \ldots \sigma_{2p}}_{\nu \rho_1 \ldots \rho_{2p}}\bar\nabla_{\sigma_1}(h'_{\sigma_2}{}^{\rho_1} n^{\rho_2})n_\mu\bar R_{\sigma_3 \sigma_4}{}^{\rho_3 \rho_4} \ldots \bar R_{\sigma_{2p-1} \sigma_{2p}}{}^{\rho_{2p-1} \rho_{2p}}+O(\omega^{-2})\;.
\end{align}
Now we can turn this expression into a total derivative.  The Bianchi identity and the identity $\bar\nabla_\mu n_\nu=\bar\nabla_\nu n_\mu$ (recall $n$ is a gradient) causes the extra terms to vanish.  The result is
\be
n_\mu E^\mu{}_\nu=\sum_{p \ge 0} \bar\nabla_{\sigma_1}\left[\Omega'p k_p \delta^{\mu \sigma_1 \ldots \sigma_{2p}}_{\nu \rho_1 \ldots \rho_{2p}}r_{\sigma_2}{}^{\rho_1} n^{\rho_2}n_\mu\bar R_{\sigma_3 \sigma_4}{}^{\rho_3 \rho_4} \ldots \bar R_{\sigma_{2p-1} \sigma_{2p}}{}^{\rho_{2p-1} \rho_{2p}}\right]+O(\omega^{-2})\;,
\ee
where we have replaced $h_{\mu\nu}$ with $\Omega r_{\mu\nu}$.  This is proportional to the derivative of the principal symbol \eqref{principalsymbol}.  Since $r_{\mu\nu}$ belongs to the kernel of the principal symbol, this quantity vanishes.

\end{document}